\documentclass[10pt, final]{IEEEtran}

\usepackage{amsthm}

\usepackage{amssymb}
\usepackage{amsmath}
\usepackage{bbm}
\usepackage{mathrsfs}
\usepackage{cite}
\usepackage{bm,epsfig,amsthm,url}
\usepackage{indentfirst}
\usepackage{amsfonts}
\usepackage{epstopdf}
\usepackage{cases}
\usepackage[caption=false,font=scriptsize]{subfig}
\usepackage{xcolor}
\usepackage{mathtools}

\makeatletter
\renewcommand{\maketag@@@}[1]{\hbox{\m@th\normalsize\normalfont#1}}%
\makeatother
\usepackage{hyperref}
\usepackage{stfloats}

\newtheoremstyle{mystyle}{}{}{}{}{}{: }{0pt}{\indent \it{\thmname{#1}\thmnumber{ #2}\thmnote{#3}}}
\theoremstyle{mystyle}

\newtheorem{Proposition}{Proposition}

\usepackage[noend]{algpseudocode}
\usepackage{float}
\usepackage{algorithmicx}
\usepackage[linesnumbered,ruled,vlined]{algorithm2e}

\allowdisplaybreaks[4]

\begin{document}

\title{\fontsize{18pt}{26pt}\selectfont Active IRS Assisted Joint Uplink and Downlink Communications}

\author{{Qiaoyan~Peng,~Qingqing~Wu,~Guangji~Chen,~Wen~Chen,~Shaodan~Ma}
	\thanks{Q. Peng is with the Department of Electronic Engineering, Shanghai Jiao Tong University, Shanghai 200240, China, and also with the State Key Laboratory of Internet of Things for Smart City and the Department of Electrical and Computer Engineering, University of Macau, Macao SAR, China (email: qiaoyan.peng@connect.um.edu.mo). 
	Q. Wu, and W. Chen are with the Department of Electronic Engineering, Shanghai Jiao Tong University, Shanghai 200240, China (qingqingwu@sjtu.edu.cn; wenchen@sjtu.edu.cn).
	G. Chen is with Nanjing University of Science and Technology, Nanjing 210094, China (email: guangjichen@njust.edu.cn).
	S. Ma is with the State Key Laboratory of Internet of Things for Smart City, University of Macau, Macao 999078, China (email: shaodanma@um.edu.mo).
	}
}

\maketitle
\begin{abstract}
In this paper, we investigate an intelligent reflecting surface (IRS) aided wireless communication system, where active IRSs (AIRSs) are deployed to assist communication between a base station (BS) and users of both the uplink (UL) and downlink (DL). We aim to maximize the weighted sum rate (WSR) of UL and DL communications through joint optimization of BS, AIRS beamforming, and AIRS element allocation. First, we study three deployment schemes, namely distributed AIRSs, BS-side AIRS, and user-side AIRS. For distributed AIRSs, both optimal and near-optimal solutions are derived in closed form. To draw useful insights, we analytically compare the deployment schemes in terms of the rate performance under the single-user setup. For the multi-user case, we consider two beamforming setups at the distributed AIRSs to balance performance and complexity tradeoffs. Regarding the user-adaptive AIRS beamforming, different AIRS beamforming vectors are adopted for each user; while for the static AIRS beamforming, all users share the same beamforming vectors, with identical phase shifts but different amplitudes for UL and DL. With the user-adaptive AIRS beamforming, we focus on the optimization of element allocation for rate maximization. With static AIRS beamforming, we solve the rate maximization problem by optimizing the BS transmit/receive beamformers, user beamforming, and AIRS beamforming. Despite its non-convexity, we develop an efficient alternating optimization (AO) based algorithm that solves each sub-problem optimally. Numerical results validate the practical advantages of distributed AIRSs compared to passive IRS (PIRS), BS-side AIRS, and user-side AIRS, and highlight the benefits of dynamic IRS beamforming.
\end{abstract}

\section{Introduction}
The sixth-generation (6G) and future wireless networks impose increasingly stringent requirements on both uplink (UL) and downlink (DL) transmissions. UL communication demands high energy efficiency and robust signal reception, while DL communication emphasizes high data rates, seamless coverage continuity, and low latency. To address these challenges, intelligent reflecting surface (IRS) has been recognized as an effective solution, offering significant advantages for both UL and DL communications and facilitating efficient joint optimization \cite{ULDL1,ULDL2}. By dynamically reconfiguring the wireless environment through the control of electromagnetic wave propagation, IRSs can effectively reshape signal paths to enhance communication performance. Due to its low power consumption and deployment flexibility, it has attracted interest from both academia and industry for theoretical research and hardware implementations. However, the passive IRS (PIRS) suffers from severe product-distance path loss. To overcome the multiplicative fading effect in PIRS-aided systems, active IRS (AIRS) has been proposed in \cite{active1,active2,active3}, which amplifies incident signals with low-cost hardware. It has driven extensive research on leveraging AIRS under various system setups, including Internet-of-Things networks \cite{chen_iot, wang_iot,Kumar_iot}, multiple-input single-output (MISO) systems \cite{zhu_MISO,Zargari_MISO,ma_MISO}, and wireless powered communication networks \cite{gao_wpt,zhai_wpt,chen_wpt}. 

Different from the PIRS, which is typically deployed right above the BS or user to reduce path loss in both UL and DL communication, the optimal AIRS placement depends critically on the communication link \cite{AIRS_magazine,active4,urllc1,urllc2,urllc3,iot1,iot2,iot3}. Specifically, for a single-user setup, a pioneering study \cite{active4} revealed that placing the AIRS closer to the receiver with a small amplification power helps balance signal and noise amplification, thereby yielding better rate performance. This is because the amplification factors of the AIRS are primarily determined by the channel gain of the first hop given amplification power constraints. When the AIRS is closer to the receiver, the second-hop channel experiences less path loss. As a result, a smaller amplification factor can be adopted at the AIRS, thereby reducing noise amplification power while maintaining a strong received signal. However, such a link-dependent placement strategy introduces a fundamental trade-off, i.e., a deployment strategy that benefits one communication link may degrade the performance in the other link. Specifically, the DL may suffer from increased path loss and noise amplification when the AIRS is placed closer to the BS to optimize UL performance, and vice versa. This trade-off is further exacerbated in scenarios with stringent requirements on both links, such as ultra-reliable low-latency communications (URLLC) \cite{urllc1,urllc2,urllc3} and Internet-of-Things (IoT) networks \cite{iot1,iot2,iot3}. Moreover, the SNR scaling order with respect to (w.r.t.) the number of AIRS elements $N$ in an AIRS-aided system is lower than that in its passive counterpart, i.e., $(\mathcal{O}(N))$ versus $(\mathcal{O}(N^2))$. This indicates that, although increasing the number of active elements can offer improvement, it can not fully address the inherent limitations associated with AIRS placement. To this end, deploying a single AIRS with traditional optimization strategies designed only for UL or DL communication may not achieve optimal performance for joint communication.

To overcome the aforementioned limitations and challenges introduced by asymmetry, a promising solution is to deploy two distributed AIRSs with one located near the BS and the other positioned close to the users. Since the BS-side AIRS and the user-side AIRS are dedicated to enhancing UL and DL communications, respectively, one practical issue arises naturally: how to determine the number of active elements allocated between the two AIRSs. Note that an improper element allocation may result in one link being significantly weaker than the other, thus undermining the potential of distributed AIRS. Therefore, the element allocation design is crucial to enhancing signal strength and reliability in both communication links, thereby improving joint communication performance. Furthermore, the element allocation strategy is closely intertwined with the design of the AIRS beam pattern \cite{peng_deployment,magazine}. Given the element allocation strategy, each AIRS  independently focuses on optimizing its phase shifts and amplification factors for its corresponding link. By jointly optimizing element allocation and beamforming design, the system can achieve a better balance between UL and DL performance, ensuring that neither link is significantly limited by an inefficient resource distribution. This is because the optimal element allocation should unleash the potential of beamforming under various channel conditions, and conversely, the beamforming design should be tailored to the available resources determined by the element allocation. Through such joint optimization, distributed AIRS systems can fully leverage their performance advantages while ensuring efficient resource utilization and maximizing the overall system rate.

Existing studies have explored dynamic IRS beamforming as a potential solution to enhance system performance \cite{dynamic,chu_ULDL,liu_ULDL}. Specifically, the IRS can be adjusted between UL and DL transmissions to reconfigure the wireless environment in real-time, thereby supporting low-latency and energy-efficient communications. Under the time division multiple access (TDMA) scheme with a single antenna at the BS, the IRS phase shifts for UL wireless information (WIT) transmission and DL wireless energy transmission (WET) were derived in closed form in \cite{chu_ULDL}. The results were further extended to the multi-antenna BS setup in \cite{liu_ULDL}, where the IRS can be deployed to maximize computation efficiency in mobile edge computing systems. However, a key challenge lies in the substantial feedback overhead and latency involved in adopting dedicated AIRS beamformers for UL and DL with frequent AIRS configuration \cite{lu_control,wu_control}. These issues become even more pronounced as the number of active elements increases. Given the fundamental performance-complexity trade-off, dynamic AIRS beamforming becomes less advantageous in practical deployments. Under the above considerations, one critical question arises: how to design a more scalable and efficient alternative AIRS beamforming scheme that flexibly balances system performance and implementation complexity? To address the challenges associated with separate AIRS configurations for UL and DL transmissions, i.e., high signaling overhead and reconfiguration latency, previous studies have explored joint optimization strategies with static/constant IRS beamforming \cite{wu_ULDL, Lee_ULDL, Abouamer_ULDL, zeng_ULDL}. Specifically, one fixed set of IRS phase shifts is developed to effectively balance the UL and DL rates, thereby achieving a compromise between system efficiency and complexity. For PIRS-aided systems, a comparative analysis of dynamic and static beamforming strategies was presented in \cite{wu_ULDL}. An alternating optimization (AO) framework was developed in \cite{Lee_ULDL} for IRS-aided single-user communications, where the BS and user precoding matrices are jointly optimized through eigenmode transmission to maximize the weighted sum rate (WSR). It was demonstrated in \cite{Abouamer_ULDL} that joint resource allocation design for multi-user MISO systems can provide significant gains over fixed UL or DL designs. Moreover, the work \cite{zeng_ULDL} investigated the integration of AIRS in wireless powered communication networks, where all devices share a common set of IRS phase shifts and amplification factors for UL WIT and DL WET subject to the imposed amplification constraint. The findings demonstrated the effectiveness of employing static AIRS beamforming to support both UL and DL communications while significantly reducing configuration overhead. Although it incurs a moderate performance loss compared to individually optimized or dynamically adjusted designs, static AIRS beamforming may offer a more practical solution in scenarios where complexity and latency are prioritized.

Driven by the aforementioned challenges, we study a joint UL and DL communication system aided by two small-sized AIRSs deployed to facilitate transmission between a BS and multiple users. Specifically, one AIRS is deployed near the BS and the other one near the users, which share the total number of elements to maximize amplification gains and minimize the negative effects of two-hop path loss. Under a time-division duplex protocol, the AIRSs take turns being activated over different time slots to support UL and DL transmissions, respectively. Based on the temporal-spatial coordination design, such a deployment not only improves the signal quality but also provides greater flexibility in adapting to dynamic channel conditions and asymmetric traffic demands. In addition, it offers particular advantages in scenarios where integrating all the active elements into a larger AIRS at a single location is impractical due to space/cost/power limitations. 
The main contributions of this paper are summarized as follows:
\begin{itemize}
\item First, we consider a special case of the line-of-sight (LoS) setup with a single user. Through theoretical analysis of three AIRS deployment architectures, namely, distributed AIRS, BS-side AIRS, and user-side AIRS, we derive closed-form expressions for WSR and establish sufficient conditions under which each architecture outperforms the others. For the distributed AIRS, we determine the optimal and near-optimal active elements at the two AIRSs and further characterize the impact of system parameters on the element allocation strategies. Our analytical results demonstrate that more elements should be deployed at the user-side AIRS when DL communication dominates or the total number of AIRS elements increases. Moreover, the distributed AIRS can achieve superior rate performance within a specific threshold weight region, which can be extended by increasing the number of BS antennas and AIRS elements, the transmit power, as well as the amplification power budget.

\item Next, motivated by the superiority of the distributed AIRS architecture, we extend our focus to the design of transmission strategies and AIRS beamforming in a multi-user setup. Specifically, we investigate two beamforming schemes, i.e., user-adaptive and static/constant. In the user-adaptive scheme, the AIRS phase shifts are individually optimized for each user during its respective UL and DL transmission slots. Given the beamforming design similar to the single-user case, we optimize the element allocation to enhance overall system performance. In contrast, the AIRS phase shifts remain constant for both UL and DL throughout the entire communication duration in the static AIRS beamforming scheme. To solve the WSR maximization problems, we propose an efficient algorithm based on semidefinite relaxation (SDR), the Lagrangian dual transform, and the Quadratic Transform techniques, which alternately update the optimization variables to achieve convergence.

\item Simulation results are presented to verify our theoretical findings and to unveil the benefits of the distributed AIRS over the BS-side AIRS, user-side AIRS, and conventional PIRS for improving the WSR under various system setups. Moreover, it demonstrates that the distributed AIRS with user-adaptive AIRS beamforming achieves significant performance gains, while with static beamforming it reduces signaling overhead and even outperforms other deployment architectures using dynamic beamforming. In addition, the static beamforming scheme with the proposed joint design can achieve a larger rate region than that of the fixed UL and fixed DL design, which demonstrates its effectiveness in balancing the UL and DL rates.
\end{itemize}

The remainder of this paper is organized as follows. Section \ref{System Model} introduces the system model of the distributed AIRS architecture. In Section \ref{Single-User System}, we provide a theoretical WSR comparison of three AIRS deployment architectures in the single-user case. Section \ref{Multi-User System} addresses the beamforming design problem for the distributed AIRS under the multi-user setup. Numerical results are presented in Section \ref{Simulation} to draw valuable insights and evaluate the performance of the proposed schemes. Finally, we conclude this paper in Section \ref{Conclusion}.

\textit{Notations:} Scalars are denoted by lower-case italic letters, vectors by bold-face lower-case, matrices by upper-case letters, and sets by calligraphic upper-case letters. $\mathbb{C}$ stands for the set of complex-valued matrices. $\operatorname{diag} \left(\cdot\right)$, $\left\| \cdot \right\|$, $\operatorname{tr} \left(\cdot\right)$, and $\left(\cdot\right)^H$ represent the diagonal matrix operator, Euclidean norm, matrix trace operator, and the conjugate transpose operator, respectively.

\section{System Model}
\label{System Model}
\begin{figure}[!t]
	\centering
	\includegraphics[width=0.85\linewidth]{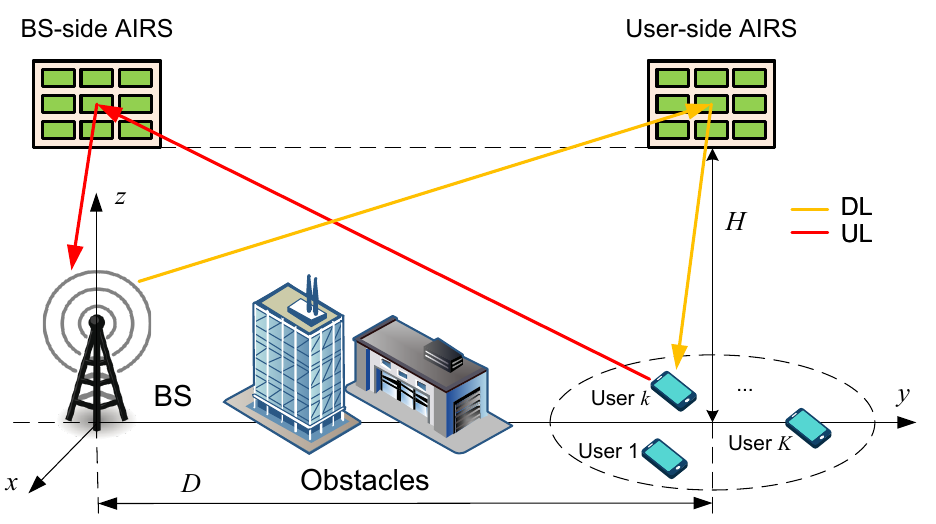}
	\caption{The distributed AIRS-aided joint uplink and downlink multi-user communication system.}
	\label{fig:multi-user}
\end{figure}

As illustrated in Fig. \ref{fig:multi-user}, we consider a joint UL and DL communication system aided by AIRS, where the BS consists of $M$ antennas and serves $K$ single-antenna users, denoted by the set $\mathcal{K} \triangleq \{1, \cdots, K\}$. Moreover, a total of $N$ AIRS elements are deployed for enhancing wireless transmissions. The active elements are divided between two distributed AIRSs, where the BS-side AIRS and user-side AIRS are equipped with $N_\mathrm{U}$ elements and $N_\mathrm{D}$ active elements, respectively, i.e., $N = N_\mathrm{U} + N_\mathrm{D}$. The distributed AIRSs operate in a time-switching manner by alternately activating one while deactivating the other, the latter achieved by setting its reflection amplitude to zero \cite{alpha}. It not only reduces overall energy consumption but also prevents the continuous operation of all active elements, which helps prolong the lifespan while maintaining the desired communication performance. We assume that the direct BS-user link is negligible due to blockage \cite{lu_deployment,hirs_peng,block_chen}. Consequently, the BS communicates with the users through single-reflection links in UL (DL), i.e., user$\to$BS-side AIRS$\to$BS (BS$\to$user-side AIRS$\to$user), while multiple-reflection paths are not considered. Moreover, the TDMA scheme with orthogonal time slots of equal duration is employed for user scheduling at the BS, adopting an equal-priority strategy that assigns identical time duration to all users. 

\subsection{Uplink}
In the $k$-th time slot of UL transmission, which is allocated to user $k$, a dedicated amplitude $\alpha _{\mathrm{U},k}$ and a dedicated phase-shift matrix ${{\bm{\Phi }} _{\mathrm{U},k}}$ are employed at the BS-side AIRS. Let ${{\bm{\Phi }} _{\mathrm{U},k}} = \operatorname{\operatorname{diag}} ( {e^{j{\phi ^{{\mathrm{U}}}_{k,1}}}}, \ldots ,{e^{j{\phi ^{{\mathrm{U}}}_{k,{N_{{\mathrm{U}}}}}}}} )$, where $\phi ^{\mathrm{U}}_{k,n}$ is the phase shift at element $n \in \mathcal{N}_\mathrm{U} \triangleq \left\{1, \cdots, N_\mathrm{U}\right\}$. Note that the AIRS amplifies not only the incoming signal but also the noise. Let $P_{\mathrm{F}}$ denote the maximum amplification power of the BS-side AIRS and thus we have
\begin{align}
	\label{power_ul}
	\alpha _{\mathrm{U},k}^2( {{p_k}{{\left\| {{\bm{\Phi }}_{\mathrm{U},k} {{\bm{h}}_{\mathrm{U},k}}} \right\|}^2} + \sigma _{\mathrm{F}}^2{{\left\| {{\bm{\Phi }} _{\mathrm{U},k}} \right\|}_F^2}} ) \le {P_{\mathrm{F}}}, \forall k \in \mathcal{K},
\end{align}
where $p_k \in [0, {P_{\mathrm{U}}}]$ denotes the transmit power of user $k$ with the maximum power $P_{\mathrm{U}}$, $\sigma _{\mathrm{F}}^2$ is the noise power at the AIRS, and $\bm{ h}_{\mathrm{U},k}$ denotes the channel from user $k$ to the BS-side AIRS.
In the $k$-th UL time slot, the received signal at the BS can be expressed as
\begin{align}
	{y}_{\mathrm{U},k} =& {\bm{u}}_k^H ( {\bm{G}}_{{\mathrm{U}}}^H{\alpha _{\mathrm{U},k}} \bm{\Phi}_{\mathrm{U},k} \bm{h}_{\mathrm{U},k} \sqrt {{p_k}} s_{\mathrm{U},k} \nonumber\\
	&+ {\bm{G}}_{{\mathrm{U}}}^H{\alpha _{\mathrm{U},k}}{\bm{\Phi } _{\mathrm{U},k}}{{\bm{n}}_{{\mathrm{U}}}} + {{\bm{n}}_0}),
\end{align}
where ${s_k^\mathrm{U}} \in \mathbb{C}$ is the transmitted signal satisfying $\mathbb{E} \{ |{s_k^\mathrm{U}}|^2 \} = 1$, ${\bm{G}}_{{\mathrm{U}}}$ denotes the channel from BS to BS-side AIRS, and ${\bm{u}}_k \in \mathbb{C}^{M \times 1}$ represents the unit-norm receive beamforming vector of the BS associated with user $k$. The amplification noise generated at the AIRS and the receive noise at the BS are represented as ${{\bm{n}}_{{\mathrm{U}}}}$ and $\bm{n}_0$, respectively, where ${{\bm{n}}_{{\mathrm{U}}}} \sim \mathcal{CN} (\bm{0}_{N_{\mathrm{U}}}, \sigma_{\mathrm{F}}^2 \bm{I}_{N_{\mathrm{U}}})$ with power $\sigma_{\mathrm{F}}^2$ and $\bm{n}_0 \sim \mathcal{CN} (\bm{0}_{M}, \sigma_0^2 \bm{I}_{M})$ with power $\sigma_0^2$. Accordingly, the achievable rate associated with user $k$ in bps/Hz can be obtained as
\begin{align}
	\label{R_Uk}
	R_{\mathrm{U},k} \!\!=\!\! \frac{1}{K}{\log _2} \!\! \left(\!\! 1 \!\!+\!\! \frac{{{p_k}{{\left| {{\bm{u}}_k^H{\bm{G}}_{{\mathrm{U}}}^H{\alpha _{\mathrm{U},k}}{{\bm{\Phi }_{\mathrm{U},k}}}{\bm{h}}_{\mathrm{U},k}} \right|}^2}}}{{{\bm{u}}_k^H ( {\alpha _{\mathrm{U},k}^2\sigma _{\mathrm{F}}^2{\bm{G}}_{{\mathrm{U}}}^H{{\bm{\Phi }_{\mathrm{U},k}}}{\bm{\Phi }_{\mathrm{U},k}^H}{{\bm{G}}_{{\mathrm{U}}}} \!\!+\!\! \sigma _0^2{{\bm{I}}_M}} \!) {{\bm{u}}_k}}} \!\! \right)\!.
\end{align}

\subsection{Downlink}
In the $k$-th time slot of DL transmission, which is allocated to user $k$, a dedicated amplitude $\alpha _{\mathrm{D},k}$ and a dedicated phase-shift matrix ${{\bm{\Phi }}_{\mathrm{D},k}}$ are employed at the user-side AIRS. Let ${{\bm{\Phi }}_{\mathrm{D},k}} = \operatorname{\operatorname{diag}} ( {e^{j{\phi ^{\mathrm{D}}_{k,1}}}}, \ldots ,{e^{j{\phi ^{\mathrm{D}}_{k,{N_{{\mathrm{D}}}}}}}} )$, where $\phi ^{\mathrm{D}}_{k,n}$ is the phase shift at element $n \in \mathcal{N}_\mathrm{D} \triangleq \left\{1, \cdots, N_\mathrm{D}\right\}$. The constraint on the amplification power of the signal reflected by the user-side AIRS is expressed as
\begin{align}
	\label{power_dl}
	\alpha _{\mathrm{D},k}^2( {{{\left\| {{{\bm{\Phi }}_{\mathrm{D},k}}{{\bm{G}}_{{\mathrm{D}}}}{\bm{w}_k}} \right\|}^2} + \sigma _{\mathrm{F}}^2{{\left\| {{{\bm{\Phi }}_{\mathrm{D},k}}} \right\|}_F^2}} ) \le {P_{\mathrm{F}}},
\end{align}
where ${{\bm{w}_k}}$ denotes the BS transmit beamforming vector satisfying ${{{\left\| {{{\bm{w}}_k}} \right\|}^2}}  \le {P_{\mathrm{B}}}$, and $\bm{G}_{{\mathrm{D}}}$ denotes the channel from BS to user-side AIRS.
The received signal at user $k$ in the DL is expressed as 
\begin{align}
	y_{\mathrm{D},k} =& {\bm{h}}_{\mathrm{D},k}^H{\alpha _{\mathrm{D},k}}{\bm{\Phi }_{\mathrm{D},k}}{{\bm{G}}_{{\mathrm{D}}}}{{\bm{w}}_k} s_{\mathrm{D},k} \nonumber\\
	&+ {\bm{h}}_{\mathrm{D},k}^H{\alpha _{\mathrm{D},k}}{\bm{\Phi }_{\mathrm{D},k}}{{\bm{n}}_{{\mathrm{D}}}} + {n_k},
\end{align}
where $\bm{h}_{\mathrm{D},k}$, ${s_{\mathrm{D},k}}$, ${{\bm{n}}_{{\mathrm{D}}}} \sim \mathcal{CN} (\bm{0}_{N_{\mathrm{D}}}, \sigma_{\mathrm{F}}^2 \bm{I}_{N_{\mathrm{D}}})$, and ${n_k} \sim \mathcal{CN} (0, \sigma_0^2)$ denote the channel from the user-side AIRS to user $k$, the transmitted signal with unit power, the amplification noise with power $\sigma_{\mathrm{F}}^2$, and the additive white Gaussian noise at the user $k$ with power $\sigma_0^2$, respectively. 
Accordingly, the achievable rate associated with user $k$ in bps/Hz is
\begin{align}
	\label{R_Dk}
	R_{\mathrm{D},k} = \frac{1}{K}{\log _2} \left( {1 + \frac{{{{| {{\bm{h}}_{\mathrm{D},k}^H{\alpha _{\mathrm{D},k}}{{\bm{\Phi }_{\mathrm{D},k}}}{{\bm{G}}_{{\mathrm{D}}}}{{\bm{w}}_k}} |}^2}}}{{\alpha _{\mathrm{D},k}^2\sigma _{\mathrm{F}}^2{{\| {{\bm{h}}_{\mathrm{D},k}^H{{\bm{\Phi }_{\mathrm{D},k}}}} \|}^2} + \sigma _0^2}}} \right).
\end{align}

Let $\varepsilon \in [0,1]$ denote the predefined weight to capture the relative priority of DL communication, with the UL weight given by $(1 - \varepsilon)$. Based on \eqref{R_Uk} and \eqref{R_Dk}, the WSR of UL and DL communication aided by two distributed AIRSs is
\begin{align}
	R_\mathrm{joint} = \sum \nolimits_{k = 1}^K \left((1-\varepsilon) R_\mathrm{U,k} +\varepsilon R_\mathrm{D,k}\right).
\end{align}

\section{Single-User System}
\label{Single-User System}
In this section, we consider the single user setup and provide a comparison among three different deployment strategies, namely the distributed AIRS, the BS-side AIRS, and the user-side AIRS. To be specific, for the BS-side AIRS, all the $N$ active elements form one single AIRS, which is placed right above the BS. For the user-side AIRS, all the $N$ active elements form one single AIRS, which is placed right above the user. For the distributed AIRS scheme, the small-sized BS-side AIRS and user-side AIRS operate in a time-switched manner by alternately activating one and deactivating the other. Let $H$ and $D$ denote the height of AIRSs and the BS-user distance, respectively.  

\begin{figure*}[ht]
	\centering
	\subfloat[Distributed AIRS]{\label{fig:distributed}\includegraphics[width=0.33\textwidth]{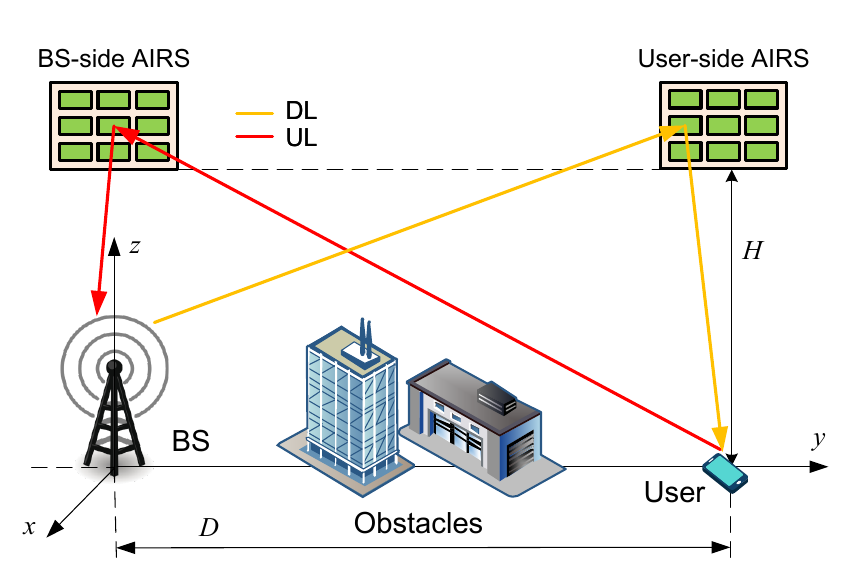}}
	\subfloat[BS-side AIRS]{\label{fig:uplink}\includegraphics[width=0.33\textwidth]{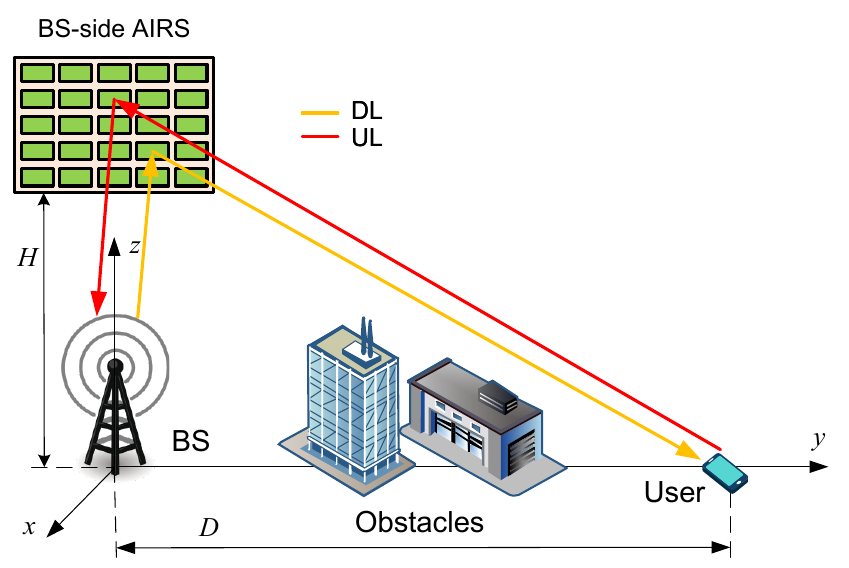}}
	\subfloat[User-side AIRS]{\label{fig:downlink}\includegraphics[width=0.33\textwidth]{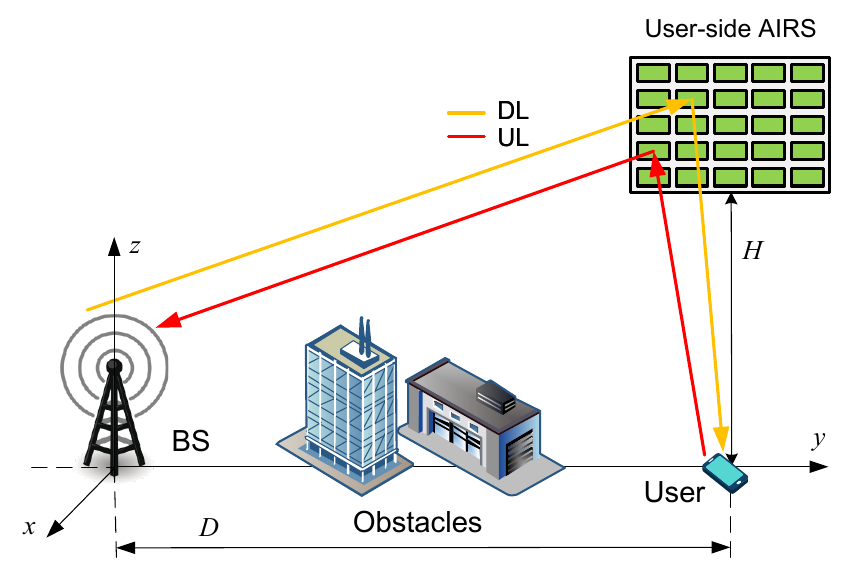}}
	\caption{Joint UL and DL communication systems aided by different AIRS deployment architectures under the single-user setup.}
	\label{fig:single-user}
\end{figure*}

\subsection{System Model}
\subsubsection{Uplink}
Let ${{\bm{a}}_{\mathrm{r}}} ( \theta _{{\mathrm{BI,UL}}}^{\mathrm{A}},\nu _{{\mathrm{BI,UL}}}^{\mathrm{A}},{N_{{\mathrm{U}}}} ) = {\bm{\mu}} ( \frac{{2{d_{\mathrm{I}}}}}{\lambda } \sin ( {\theta _{{\mathrm{BI,UL}}}^{\mathrm{A}}} )\sin ( {\nu _{{\mathrm{BI,UL}}}^{\mathrm{A}}} ),{N_{{\mathrm{UL,h}}}} ) \otimes {\bm{\mu}} ( \frac{{2{d_{\mathrm{I}}}}}{\lambda }\cos  ( {\nu _{{\mathrm{BI,UL}}}^{\mathrm{A}}} ),{N_{{\mathrm{UL,v}}}} )$ denote the receive steering vector, where $d_\mathrm{I}$ represents the inter-element spacing, $\lambda$ stands for the wavelength, ${\theta _{{\mathrm{BI,UL}}}^{\mathrm{A}}} ({\nu _{{\mathrm{BI,UL}}}^{\mathrm{A}}})$ is the azimuth (elevation) angle-of-arrival at the BS-side AIRS from BS, and ${N_{{\mathrm{U}}}} = {N_{{\mathrm{UL,h}}}}{N_{{\mathrm{UL,v}}}}$ represents the total number of active elements with the number of horizontal (vertical) elements ${N_{{\mathrm{UL,h}}}} ({N_{{\mathrm{UL,v}}}})$. The receive steering vector ${{\bm{a }}_{\mathrm{r}}} ( \theta _{{\mathrm{IU,UL}}}^{\mathrm{A}},\nu _{{\mathrm{IU,UL}}}^{\mathrm{A}}, N_{\mathrm{U}} )$ and the transmit steering vector ${{\bm{a }}_{\mathrm{t}}} ( {\theta _{{\mathrm{BI,UL}}}^{\mathrm{D}},\nu _{{\mathrm{BI,UL}}}^{\mathrm{D}},M} )$ can be defined similarly to ${{\bm{a }}_{\mathrm{r}}} ( {\theta _{{\mathrm{BI,UL}}}^{\mathrm{A}},\nu _{{\mathrm{BI,UL}}}^{\mathrm{A}},{N_{{\mathrm{U}}}}} )$. As such, the BS-AIRS channel is modeled as ${{\bm{G}}_{{\mathrm{U}}}} = {h_{{\mathrm{BI,UL}}}}{{\bm{a }}_{\mathrm{r}}} ( {\theta _{{\mathrm{BI,UL}}}^{\mathrm{A}},\nu _{{\mathrm{BI,UL}}}^{\mathrm{A}},{N_{{\mathrm{U}}}}} ){\bm{a }}_{\mathrm{t}}^H ( {\theta _{{\mathrm{BI,UL}}}^{\mathrm{D}},\nu _{{\mathrm{BI,UL}}}^{\mathrm{D}},M} )$ with the channel gain $|h_{{\mathrm{BI,UL}}}|^2$. The AIRS-user channel is modeled as $\bm{h}_{{\mathrm{U}}} = {h_{{\mathrm{IU,UL}}}} {{\bm{a }}_{\mathrm{r}}} ( \theta _{{\mathrm{IU,UL}}}^{\mathrm{A}},\nu _{{\mathrm{IU,UL}}}^{\mathrm{A}}, N_{\mathrm{U}} )$ with the channel gain $|{h_{{\mathrm{IU,UL}}}|^2}$. Thus, the received signal at the BS is
\begin{align}
	y_{{\mathrm{U}}} = {\bm{u}}^H ({\bm{G}}_{{\mathrm{U}}}^H{\alpha _{{\mathrm{U}}}}{\bm{\Phi }}_{{\mathrm{U}}} ( {{\bm{h}}_{{\mathrm{U}}}{\sqrt{p_1}}{s_\mathrm{U}} + {{\bm{n}}_{{\mathrm{U}}}}} ) + {\bm{n}_0}),
\end{align}
where $p_1 \in \left[0, P_\mathrm{U}\right]$ denotes the transmit power, ${s_\mathrm{U}} \in \mathbb{C}$ is the transmitted signal with unit power, ${\bm{\Phi }}_{{\mathrm{U}}}$ represents the phase-shift matrix, $\alpha_{{\mathrm{U}}}$ is the amplification factor of the BS-side AIRS, and $\bm{u} \in \mathbb{C}^{M \times 1} $ denotes the unit-norm receive beamforming vector of the BS. Therefore, the DL achievable rate in bps/Hz is given by
\begin{align}
	\label{RU_single}
	R_{\mathrm{U}} \!=\! {\log _2} \! \left( \! 1 \!+\! \frac{{{p_1}{{\left| {\bm{u}^H{\bm{G}}_{{\mathrm{U}}}^H{\alpha _{\mathrm{U}}}{{\bm{\Phi }_{\mathrm{U}}}}{\bm{h}}_{\mathrm{U}}} \right|}^2}}}{{{\bm{u}}^H ( {\alpha _{\mathrm{U}}^2\sigma _{\mathrm{F}}^2{\bm{G}}_{{\mathrm{U}}}^H{{\bm{\Phi }_{\mathrm{U}}}}{\bm{\Phi }_{\mathrm{U}}^H}{{\bm{G}}_{{\mathrm{U}}}} \!+\! \sigma _0^2{{\bm{I}}_M}} ) {{\bm{u}}}}} \! \right).
\end{align}
For UL communication, the AIRS reflects the signal towards the BS. The optimal AIRS phase shifts should align the cascaded BS-IRS-user channel, i.e., ${\left[ {{\bm{\Phi }}_{{\mathrm{U}}}^{\mathrm{opt}}} \right]_n} = e^{j(- \arg ( {{{\left[ {{\bm{h}}_{{\mathrm{U}}}} \right]}_n}} ) + \arg ( {{{\left[ {{{\bm{a }}_{\mathrm{r}}}( {\theta _{{\mathrm{BI,UL}}}^{\mathrm{A}},\nu _{{\mathrm{BI,UL}}}^{\mathrm{A}},{N_{{\mathrm{U}}}}} )} \right]}_n}} ))}$. The optimal amplification factors can be obtained by taking the equality of the power constraint \eqref{power_ul}, i.e., $\alpha _{{\mathrm{U}}}^{\mathrm{opt}} = \sqrt {\frac{{{P_{\mathrm{F}}}}}{{{N_{{\mathrm{U}}}}( {{p_1}h_{{\mathrm{IU,UL}}}^2 + \sigma _{\mathrm{F}}^2} )}}}$. By substituting $\alpha _{{\mathrm{U}}}^{\mathrm{opt}}$ into \eqref{RU_single}, it is readily verified that $R_{\mathrm{U}}$ monotonically increases with $p_1$. Thus, we have $p_1^{\mathrm{opt}} = P_\mathrm{U}$. Under the LoS channel, the optimal receive beamforming corresponds to the maximum ratio combining (MRC), i.e., ${\bm{u}}^{\mathrm{opt}} = \frac{{{\bm{a }}_{\mathrm{t}} ( {\theta _{{\mathrm{BI,UL}}}^{\mathrm{D}},\nu _{{\mathrm{BI,UL}}}^{\mathrm{D}},M} )}}{{\left\| {{\bm{a }}_{\mathrm{t}} ( {\theta _{{\mathrm{BI,UL}}}^{\mathrm{D}},\nu _{{\mathrm{BI,UL}}}^{\mathrm{D}},M} )} \right\|}}$. With the optimized beamforming design, the UL achievable rate in bps/Hz is given by
\begin{align}
	\label{R_UL}
	R_{{\mathrm{U}}} \!=\!  {\log _2} \! \left( \! 1  \!+\! \frac{{{P_{\mathrm{U}}}{P_{\mathrm{F}}}MN_{{\mathrm{U}}}^2h_{{\mathrm{IU,UL}}}h_{{\mathrm{BI,UL}}}^2}}{{  { M{P_{\mathrm{F}}} h_{{\mathrm{BI,UL}}}^2\sigma _{\mathrm{F}}^2 \!+\!  {P_{\mathrm{U}}}h_{{\mathrm{IU,UL}}}^2\sigma _0^2  \!+\!  \sigma _{\mathrm{F}}^2\sigma _0^2} }}  \right).
\end{align}

\subsubsection{Downlink}
The receive steering vector ${{\bm{a }}_{\mathrm{r}}}( {\theta _{{\mathrm{BI,DL}}}^{\mathrm{A}},\nu _{{\mathrm{BI,DL}}}^{\mathrm{A}},{N_{{\mathrm{D}}}}} )$, ${{\bm{a }}_{\mathrm{r}}}( {\theta _{{\mathrm{IU,DL}}}^{\mathrm{A}},\nu _{{\mathrm{IU,DL}}}^{\mathrm{A}},{N_{{\mathrm{D}}}}} )$ and the transmit steering vector ${\bm{a }}_{\mathrm{t}}( {\theta _{{\mathrm{BI,DL}}}^{\mathrm{D}},\nu _{{\mathrm{BI,DL}}}^{\mathrm{D}},M} )$ can be defined similarly to ${{\bm{a }}_{\mathrm{r}}} ( {\theta _{{\mathrm{BI,UL}}}^{\mathrm{A}},\nu _{{\mathrm{BI,UL}}}^{\mathrm{A}},{N_{{\mathrm{U}}}}} )$. As such, the channel between BS and user-side AIRS can be modeled as ${{\bm{G}}_{{\mathrm{D}}}} = {h_{{\mathrm{BI,DL}}}}{{\bm{a }}_{\mathrm{r}}}( {\theta _{{\mathrm{BI,DL}}}^{\mathrm{A}},\nu _{{\mathrm{BI,DL}}}^{\mathrm{A}},{N_{{\mathrm{D}}}}} ){\bm{a }}_{\mathrm{t}}^H( {\theta _{{\mathrm{BI,DL}}}^{\mathrm{D}},\nu _{{\mathrm{BI,DL}}}^{\mathrm{D}},M})$ with the channel gain $|h_{{\mathrm{BI,DL}}}|^2$.
The AIRS-user channel is modeled as ${\bm{h}}_{{\mathrm{D}}} = {h_{{\mathrm{IU,DL}}}}{{\bm{a }}_{\mathrm{r}}}( {\theta _{{\mathrm{IU,DL}}}^{\mathrm{A}},\nu _{{\mathrm{IU,DL}}}^{\mathrm{A}},{N_{{\mathrm{D}}}}} )$ with the channel gain ${|h_{{\mathrm{IU,DL}}}|^2}$.
Similar to the UL communication model, the received signal at the user for DL communication can be expressed as 
\begin{align}
	y_{{\mathrm{D}}} = {\bm{h}}_{\mathrm{D}}^H{\alpha _{{\mathrm{D}}}}{{\bm{\Phi }}_{{\mathrm{D}}}}( {{{\bm{G}}_{{\mathrm{D}}}}{{\bm{w}}}{s_\mathrm{D}} + {{\bm{n}}_{{\mathrm{D}}}}} ) + {z_\mathrm{D}},
\end{align}
where ${z_\mathrm{D}} \sim \mathcal{CN} (0, \sigma_0^2)$ denotes the received noise at the user. Therefore, the DL achievable rate in bps/Hz is 
\begin{align}
	R_{{\mathrm{D}}} = {\log _2} \left( {1 + \frac{{{{\left| {{{\bm{h}}_{{\mathrm{D}}} ^H}{\alpha _{{\mathrm{D}}}}{{\bm{\Phi }}_{{\mathrm{D}}}}{{\bm{G}}_{{\mathrm{D}}}}{{\bm{w}}}} \right|}^2}}}{{{{\left\| {{{\bm{h}}_{{\mathrm{D}}}^H}{\alpha _{{\mathrm{D}}}}{{\bm{\Phi }}_{{\mathrm{D}}}}} \right\|}^2}\sigma _{\mathrm{F}}^2 + \sigma _0^2}}} \right).
\end{align}
Note that the BS transmits the signal towards the AIRS for DL communication. Similar to the UL communication, the optimal beamforming of the AIRS is given by ${ [ {{\bm{\Phi }}_{{\mathrm{D}}}^{\mathrm{opt}}} ]_n} = e^{j(\arg ( {{[ {{\bm{h}}_{{\mathrm{D}}}} ]_n}} ) - \arg ( {{ [ {{{\bm{a }}_{\mathrm{r}}}( {\theta _{{\mathrm{BI,DL}}}^{\mathrm{A}},\nu _{{\mathrm{BI,DL}}}^{\mathrm{A}},{N_{{\mathrm{D}}}}} )} ]_n}} )}$, $\forall n \in \mathcal{N}_\mathrm{D} \triangleq \left\{1, \cdots, {N}_\mathrm{D} \right\}$ and $\alpha _{{\mathrm{D}}}^{\mathrm{opt}} = \sqrt {\frac{{{P_{\mathrm{F}}}}}{{{N_{{\mathrm{D}}}}( {M{P_{\mathrm{B}}}h_{{\mathrm{BI,DL}}}^2 + \sigma _{\mathrm{F}}^2} )}}}$. We adopt the maximum ratio transmission (MRT) beamformer, i.e., ${\bm{w}}^{\mathrm{opt}} = \frac{ \sqrt{P_\mathrm{B}}{{\bm{a }}_{\mathrm{t}} ( {\theta _{{\mathrm{BI,DL}}}^{\mathrm{D}},\nu _{{\mathrm{BI,DL}}}^{\mathrm{D}},M} )}}{{\left\| {{\bm{a }}_{\mathrm{t}} ( {\theta _{{\mathrm{BI,DL}}}^{\mathrm{D}},\nu _{{\mathrm{BI,DL}}}^{\mathrm{D}},M} )} \right\|}}$. With the optimized beamforming design, the DL achievable rate in bps/Hz is
\begin{align}
	\label{R_DL}
	R_{{\mathrm{D}}}  \!=\!  {\log _2} \! \left( \! 1 \! + \! \frac{{{P_{\mathrm{B}}}{P_{\mathrm{F}}}MN_{{\mathrm{D}}}h_{{\mathrm{IU,DL}}}^2h_{{\mathrm{BI,DL}}}^2}}{{  {{P_{\mathrm{F}}}h_{{\mathrm{IU,DL}}}^2\sigma _{\mathrm{F}}^2  \!+\!  M{P_{\mathrm{B}}}h_{{\mathrm{BI,DL}}}^2\sigma _0^2  \!+\!  \sigma _{\mathrm{F}}^2\sigma _0^2} }} \right).
\end{align}

\subsection{Problem Formulation}
\subsubsection{The Distributed AIRSs}
We focus on the element allocation design of the distributed AIRSs. For notational simplicity, we define $h_1^2 \triangleq \beta / {H^2}$ and $h_2^2 \triangleq \beta / (D^2+H^2)$, where $\beta$ stands for the channel power gain at the reference distance of 1 meter (m). Thus, we have $|h_{{\mathrm{BI,UL}}}|^2 = |h_{{\mathrm{IU,DL}}}|^2 = h_1^2$ and $ |h_{{\mathrm{BI,DL}}}|^2 = |h_{{\mathrm{IU,UL}}}|^2 = h_2^2$. Therefore, the WSR is given by 
\begin{align}
	\label{R_DIS}
	R_\mathrm{DIS} ({N_{{\mathrm{U}}}},{N_{{\mathrm{D}}}}) =& (1-\varepsilon) {\log _2}( {1 + {{{P_{\mathrm{U}}}{f_1} { {N_{{\mathrm{U}}}}} }}/{{{f_2}}}} ) \nonumber\\
	&+ \varepsilon {\log _2}( {1 + {{{P_{\mathrm{B}}}{f_1}{N_{{\mathrm{D}}}}}}/{{{f_3}}}} ),
\end{align}
where ${f_1} = M{P_{\mathrm{F}}}h_1^2h_2^2$, ${f_2} = M {P_{\mathrm{F}}} h_1^2\sigma _{\mathrm{F}}^2 + {P_{\mathrm{U}}}h_2^2\sigma _0^2 + \sigma _{\mathrm{F}}^2\sigma _0^2$, and ${f_3} = {P_{\mathrm{F}}}h_1^2\sigma _{\mathrm{F}}^2 + M{P_{\mathrm{B}}}h_2^2\sigma _0^2 + \sigma _{\mathrm{F}}^2\sigma _0^2$. Note that element allocation optimization are based on long-term statistical channel information, while beamforming is performed by using instantaneous channel state information. Moreover, we assume that the channel state information for all the channels is available. In the following, we focus on the WSR maximization problem in an offline manner to determine the element allocation in the distributed AIRS-aided communication system in a quasi-static scenario.
Based on \eqref{R_DIS}, the problem is formulated as
\begin{subequations}
	\label{pro_single} 
	\begin{align}
		\mathop {\max }\limits_{{N_{{\mathrm{U}}}}, {N_{{\mathrm{D}}}}} \;\;\; &R_\mathrm{DIS} ({N_{{\mathrm{U}}}},{N_{{\mathrm{D}}}}) \label{y_D} \\ 
		\mathrm{s.t.} \;\;\;\;\;\;
		& {N_{{\mathrm{U}}}} + {N_{{\mathrm{D}}}} \le N \label{No.}, {N_{{\mathrm{U}}}} \in \mathbb{N}^+, {N_{{\mathrm{D}}}} \in \mathbb{N}^+.
	\end{align}
\end{subequations}
It can be proved by contradiction that constraint \eqref{No.} holds with equality at the optimal solution to problem \eqref{pro_single}, i.e., ${N_{{\mathrm{U}}}} + {N_{{\mathrm{D}}}} = N$. By relaxing the integer values ${N_{{\mathrm{U}}}}$ and ${N_{{\mathrm{D}}}}$ into their continuous counterparts, i.e., ${x_{{\mathrm{U}}}}$ and ${x_{{\mathrm{D}}}}$, we can represent $R_\mathrm{DIS} ({N_{{\mathrm{U}}}},{N_{{\mathrm{D}}}})$ in \eqref{y_D} as $R_\mathrm{DIS} ({x_{{\mathrm{D}}}}) = (1-\varepsilon ) {\log _2}( {1 + {{{P_{\mathrm{U}}}{f_1} { (N - {x_{{\mathrm{D}}}})} }}/{{{f_2}}}} ) + \varepsilon {\log _2}( {1 + {{{P_{\mathrm{B}}}{f_1}{x_{{\mathrm{D}}}}}}/{{{f_3}}}} )$. Thus, problem \eqref{pro_single} is reduced to
\begin{subequations}
	\label{pro_single_x}
	\begin{align}
		\mathop {\max }\limits_{{x_{{\mathrm{D}}}}}& \;\;\; R_\mathrm{DIS} ({x_{{\mathrm{D}}}}) \\
		\mathrm{s.t.}& \;\;\;\; 0 \le {x_{{\mathrm{D}}}} \le N.
	\end{align}
\end{subequations}
	
\begin{figure*}[ht]
	\begin{align}
		\label{x_DL}
		{x_{{\mathrm{D}}}^\mathrm{opt}} = \begin{cases}
			0,&0  \le  \varepsilon \le  \frac{{{P_{\mathrm{U}}}{f_3}}}{{{P_{\mathrm{B}}}{P_{\mathrm{U}}}{f_1}N + {P_{\mathrm{B}}}{f_2} + {P_{\mathrm{U}}}{f_3}}}, \\
			\frac{{\varepsilon ( {P_{\mathrm{B}}}{P_{\mathrm{U}}}{f_1}N + {P_{\mathrm{B}}}{f_2} + {P_{\mathrm{U}}}{f_3}) - {P_{\mathrm{U}}}{f_3}}}{{{P_{\mathrm{B}}}{P_{\mathrm{U}}}{f_1}}},&\frac{{{P_{\mathrm{U}}}{f_3}}}{{{P_{\mathrm{B}}}{P_{\mathrm{U}}}{f_1}N + {P_{\mathrm{B}}}{f_2} + {P_{\mathrm{U}}}{f_3}}}  < \varepsilon <  \frac{{{P_{\mathrm{B}}}{P_{\mathrm{U}}}{f_1}N + {P_{\mathrm{U}}}{f_3} }}{{{P_{\mathrm{B}}}{P_{\mathrm{U}}}{f_1}N + {P_{\mathrm{B}}}{f_2} + {P_{\mathrm{U}}}{f_3}}}, \\
			N,&\frac{{{P_{\mathrm{B}}}{P_{\mathrm{U}}}{f_1}N + {P_{\mathrm{U}}}{f_3}}}{{{P_{\mathrm{B}}}{P_{\mathrm{U}}}{f_1}N + {P_{\mathrm{B}}}{f_2} + {P_{\mathrm{U}}}{f_3}}} \le \varepsilon \le 1.
		\end{cases}
	\end{align}
	{\noindent} \rule[-0pt]{18.3cm}{0.05em}
\end{figure*}
\begin{Proposition}
	\label{pro_x}
	The optimal solution to problem \eqref{pro_single_x} is presented in \eqref{x_DL} on the top of next page.
\end{Proposition}

{\it{Proof:}}
The first-order partial derivative of $R_\mathrm{DIS} ({x_{{\mathrm{D}}}})$ w.r.t. ${{x_{{\mathrm{D}}}}}$ is given by $\frac{{\partial R_\mathrm{DIS} ({x_{{\mathrm{D}}}})}}{{\partial {x_{{\mathrm{D}}}}}} = \frac{{{f_1}p ( {{x_{{\mathrm{D}}}}} )}}{{\ln 2 ( {{f_3} + {P_{\mathrm{B}}}{f_1}{x_{{\mathrm{D}}}}} ) ( {{f_2} + {P_{\mathrm{U}}}{f_1} ( {N - {x_{{\mathrm{D}}}}} )} )}}$, where $q_1( {{x_{{\mathrm{D}}}}} ) =  - {P_{\mathrm{B}}}{P_{\mathrm{U}}}{f_1}{x_{{\mathrm{D}}}} + \varepsilon ( {P_{\mathrm{B}}}{P_{\mathrm{U}}}{f_1}N + {P_{\mathrm{B}}}{f_2} + {P_{\mathrm{U}}}{f_3}) - {P_{\mathrm{U}}}{f_3}$. Then, it follows that $q_1( {{x_{{\mathrm{D}}}}} )$ monotonically decreases with ${x_{{\mathrm{D}}}}$. We represent $q_1( {{x_{{\mathrm{D}}}}} = 0 )$ and $q_1( {{x_{{\mathrm{D}}}}} = N )$ as ${q_2}( \varepsilon ) = \varepsilon ( {P_{\mathrm{B}}}{P_{\mathrm{U}}}{f_1}N + {P_{\mathrm{B}}}{f_2} + {P_{\mathrm{U}}}{f_3}) - {P_{\mathrm{U}}}{f_3}$ and ${q_3}( \varepsilon ) =  - {P_{\mathrm{B}}}{P_{\mathrm{U}}}{f_1}N + \varepsilon ( {P_{\mathrm{B}}}{P_{\mathrm{U}}}{f_1}N + {P_{\mathrm{B}}}{f_2} + {P_{\mathrm{U}}}{f_3}) - {P_{\mathrm{U}}}{f_3}$, respectively. We define ${W_1} \triangleq \frac{{{P_{\mathrm{U}}}{f_3}}}{{{P_{\mathrm{B}}}{P_{\mathrm{U}}}{f_1}N + {P_{\mathrm{B}}}{f_2} + {P_{\mathrm{U}}}{f_3}}}$ and $W_2 \triangleq \frac{{{P_{\mathrm{B}}}{P_{\mathrm{U}}}{f_1}N + {P_{\mathrm{U}}}{f_3} }}{{{P_{\mathrm{B}}}{P_{\mathrm{U}}}{f_1}N + {P_{\mathrm{B}}}{f_2} + {P_{\mathrm{U}}}{f_3}}}$. Since ${q_2}( \varepsilon ) \le 0$ when $0 \le \varepsilon \le {W_1}$, we have $q_1( {{x_{{\mathrm{D}}}}} ) \le 0$ and $\frac{{\partial R_\mathrm{DIS} ({x_{{\mathrm{D}}}})}}{{\partial {x_{{\mathrm{D}}}}}} \le 0$, i.e., $R_\mathrm{DIS} ({x_{{\mathrm{D}}}})$ monotonically decreases with ${{x_{{\mathrm{D}}}}}$. Since ${q_3}( \varepsilon ) \ge 0$ when $W_2 \le \varepsilon \le 1$, we have $q_1( {{x_{{\mathrm{D}}}}} ) \ge 0$ and $\frac{{\partial R_\mathrm{DIS} ({x_{{\mathrm{D}}}})}}{{\partial {x_{{\mathrm{D}}}}}} \ge 0$, i.e., $R_\mathrm{DIS} ({x_{{\mathrm{D}}}})$ monotonically increases with ${{x_{{\mathrm{D}}}}}$. When $W_1 < \varepsilon < W_2$, there is one and only one root for $q_1( {{x_{{\mathrm{D}}}}} ) = 0$, which is ${x_{{\mathrm{D}}}^\mathrm{opt}} = \frac{{\varepsilon ( {P_{\mathrm{B}}}{P_{\mathrm{U}}}{f_1}N + {P_{\mathrm{B}}}{f_2} + {P_{\mathrm{U}}}{f_3}) - {P_{\mathrm{U}}}{f_3}}}{{{P_{\mathrm{B}}}{P_{\mathrm{U}}}{f_1}}}$.
As such, the proof is completed.
~$\hfill\blacksquare$

From proposition \ref{pro_x}, it can be readily verified that ${x_{{\mathrm{D}}}^\mathrm{opt}}$ monotonically increases with $P_\mathrm{B}$ and decreases with $P_\mathrm{U}$ when $W_1 < \varepsilon < W_2$. As such, with a larger $P_\mathrm{B}$ and/or a smaller $P_\mathrm{U}$, more elements should be deployed at the AIRS for DL communication to increase the signal power received at the user, thereby improving the WSR. By solving problem \eqref{pro_single_x}, we can apply the integer rounding technique to reconstruct the optimal solution to the original optimization problem \eqref{pro_single} based on proposition \ref{pro_x} and then obtain the optimal element allocation at the distributed AIRSs as detailed below. Therefore, the maximum WSR is given by
\begin{align}
	\label{R_D_opt}
	R _{\mathrm{DIS}}^{{\mathrm{opt}}} = \mathop {\max }\limits_{{x_{{\mathrm{D}}}}} {R _{\mathrm{DIS}}}( {{x_{{\mathrm{D}}}}} ),{x_{{\mathrm{D}}}} \in \left\{ {\left\lfloor {x_{{\mathrm{D}}}^\mathrm{opt}} \right\rfloor ,\left\lceil {x_{{\mathrm{D}}}^\mathrm{opt}} \right\rceil } \right\}.
\end{align}

To gain more useful insights, we focus on the high SNR case for both UL and DL, i.e., ${{{P_{\mathrm{U}}}{f_1} { {N_{{\mathrm{U}}}}} }}/{{{f_2}}} \gg 1$ and ${{{{P_{\mathrm{B}}}{f_1}{N_{{\mathrm{D}}}}}}/{{{f_3}}}} \gg 1$. Accordingly, ${R _{\mathrm{DIS}}}( {{x_{{\mathrm{D}}}}} )$ in problem \eqref{pro_single_x} can be approximated as ${\bar R _{\mathrm{DIS}}}( {{x_{{\mathrm{D}}}}} ) = (1-\varepsilon ){\log _2}( {{{{P_{\mathrm{U}}}{f_1}( {N - {x_{{\mathrm{D}}}}} )}}/{{{f_2}}}} ) + \varepsilon {\log _2}( {{{{P_{\mathrm{B}}}{f_1}{x_{{\mathrm{D}}}}}}/{{{f_3}}}} )$. Therefore, problem \eqref{pro_single_x} can be reformulated as
\begin{subequations}
	\label{pro_single_x_snr}
	\begin{align}
		\mathop {\max }\limits_{{x_{{\mathrm{D}}}}}& \;\;\; \bar R_\mathrm{DIS} ({x_{{\mathrm{D}}}}) \\
		\mathrm{s.t.}& \;\;\;\; 0 < {x_{{\mathrm{D}}}} < N.
	\end{align}
\end{subequations}
Under the assumption of ${{{P_{\mathrm{U}}}{f_1} { {N_{{\mathrm{U}}}}} }}/{{{f_2}}} \gg 1$ and ${{{{P_{\mathrm{B}}}{f_1}{N_{{\mathrm{D}}}}}}/{{{f_3}}}} \gg 1$, we derive a near-optimal solution for element allocation, which is much more tractable and suitable for practical implementation as detailed below.
\begin{Proposition}
	\label{pro:x}
	The optimal solution to problem \eqref{pro_single_x} is
	\begin{align}
		\label{x_DL_subopt}
		{x_{{\mathrm{D}}}^\mathrm{near-opt}} = \varepsilon N, 0 < \varepsilon < 1.
	\end{align}
\end{Proposition}

{\it{Proof:}}
The first-order partial derivative of $\bar R_\mathrm{DIS} ({x_{{\mathrm{D}}}})$ w.r.t. ${{x_{{\mathrm{D}}}}}$ is given by $\frac{{\partial \bar R ({x_{{\mathrm{D}}}})}}{{\partial {x_{{\mathrm{D}}}}}} = \frac{{ \varepsilon N - {x_{{\mathrm{D}}}}}}{{\ln 2{x_{{\mathrm{D}}}}( {N - {x_{{\mathrm{D}}}}} )}}$. When ${x_{{\mathrm{D}}}} \in ( {0, \varepsilon N} )$, it follows that $\frac{{\partial \bar R ({x_{{\mathrm{D}}}})}}{{\partial {x_{{\mathrm{D}}}}}} > 0$, i.e., $\bar R_\mathrm{DIS} ({x_{{\mathrm{D}}}})$ monotonically increases with ${{x_{{\mathrm{D}}}}}$. When ${x_{{\mathrm{D}}}} \in ( {\varepsilon N, N} )$, it follows that $\frac{{\partial \bar R ({x_{{\mathrm{D}}}})}}{{\partial {x_{{\mathrm{D}}}}}} < 0$, i.e., $\bar R_\mathrm{DIS} ({x_{{\mathrm{D}}}})$ monotonically decreases with ${{x_{{\mathrm{D}}}}}$. There is one and only one root for $\frac{{\partial \bar R ({x_{{\mathrm{D}}}})}}{{\partial {x_{{\mathrm{D}}}}}} = 0$, which is ${x_{{\mathrm{D}}}^\mathrm{near-opt}} = \varepsilon N$. The proof is thus completed.
~$\hfill\blacksquare$

Proposition \eqref{pro:x} provides a straightforward and effective guideline for element allocation, indicating that the number of active elements allocated at the user-side AIRS is proportional to the weight for DL communication $\varepsilon$ and the total number of AIRS elements $N$. Given $N$, more elements should be allocated to the user-side AIRS for DL communication as $\varepsilon$ increases, thereby maximizing the WSR. This is expected because a higher weight for DL communication indicates greater importance for DL, and thus should be matched by a corresponding increase in element allocation to fully exploit the potential of the system.  

\subsubsection{The BS-side AIRS}
We next consider the case where a single AIRS is deployed directly above the BS for both UL and DL communication, which implies $|h_{{\mathrm{BI,UL}}}|^2 = |h_{{\mathrm{BI,DL}}}|^2 = h_1^2$, $|h_{{\mathrm{IU,UL}}}|^2 = |h_{{\mathrm{IU,DL}}}|^2 = h_2^2$, and $N_\mathrm{U} = N_\mathrm{D} = N$. Based on \eqref{R_UL} and \eqref{R_DL}, the corresponding WSR is given by
\begin{align}
	\label{y_B}
	R_{\mathrm{BS}} = & (1-\varepsilon ){\log _2}  \left( {1 + \frac{{{P_{\mathrm{U}}}{P_{\mathrm{F}}}M N h_1^2h_2^2}}{{M{P_{\mathrm{F}}}h_1^2\sigma _{\mathrm{F}}^2 + {P_{\mathrm{U}}}h_2^2\sigma _0^2 + \sigma _{\mathrm{F}}^2\sigma _0^2}}}  \right) \nonumber\\ 
	+& \varepsilon {\log _2} \left( {1 + \frac{{{P_{\mathrm{B}}}{P_{\mathrm{F}}}MNh_1^2h_2^2}}{{{P_{\mathrm{F}}}h_2^2\sigma _{\mathrm{F}}^2 + M{P_{\mathrm{B}}}h_1^2\sigma _0^2 + \sigma _{\mathrm{F}}^2\sigma _0^2}}}  \right).
\end{align}

\subsubsection{The user-side AIRS}
Finally, we study the case where a single AIRS with $N$ elements is positioned above the user for both UL and DL communication. Then, it follows that $h_{{\mathrm{BI,UL}}}^2 = h_{{\mathrm{BI,DL}}}^2 = h_2^2$, $h_{{\mathrm{IU,UL}}}^2 = h_{{\mathrm{IU,DL}}}^2 = h_1^2$, and $N_\mathrm{U} = N_\mathrm{D} = N$. Based on \eqref{R_UL} and \eqref{R_DL}, the WSR is given by
\begin{align}
	\label{y_U}
	R_{\mathrm{UE}} = & (1-\varepsilon ){\log _2}  \left( {1 + \frac{{{P_{\mathrm{U}}}{P_{\mathrm{F}}}MNh_1^2h_2^2N}}{{M{P_{\mathrm{F}}}h_2^2\sigma _{\mathrm{F}}^2 + {P_{\mathrm{U}}}h_1^2\sigma _0^2 + \sigma _{\mathrm{F}}^2\sigma _0^2}}}  \right) \nonumber\\ 
	+& \varepsilon {\log _2}  \left(  {1 + \frac{{{P_{\mathrm{B}}}{P_{\mathrm{F}}}MNh_1^2h_2^2}}{{{P_{\mathrm{F}}}h_1^2\sigma _{\mathrm{F}}^2 + M{P_{\mathrm{B}}}h_2^2\sigma _0^2 + \sigma _{\mathrm{F}}^2\sigma _0^2}}}  \right).
\end{align}

\subsubsection{Comparison}
\begin{figure*}[ht]
	\begin{align}
		\label{y_opt}
		R^{\mathrm{opt}} = \begin{cases}
			\max (R_{\mathrm{BS}},R_{\mathrm{UE}}),&0  \le  \varepsilon \le  \frac{{{P_{\mathrm{U}}}{f_3}}}{{{P_{\mathrm{B}}}{P_{\mathrm{U}}}{f_1}N + {P_{\mathrm{B}}}{f_2} + {P_{\mathrm{U}}}{f_3}}}, \\
			\max (R_{\mathrm{BS}},R_{\mathrm{UE}},R_{\mathrm{DIS}}^{\mathrm{opt}} ),&\frac{{{P_{\mathrm{U}}}{f_3}}}{{{P_{\mathrm{B}}}{P_{\mathrm{U}}}{f_1}N + {P_{\mathrm{B}}}{f_2} + {P_{\mathrm{U}}}{f_3}}}  < \varepsilon <  \frac{{{P_{\mathrm{B}}}{P_{\mathrm{U}}}{f_1}N + {P_{\mathrm{U}}}{f_3} }}{{{P_{\mathrm{B}}}{P_{\mathrm{U}}}{f_1}N + {P_{\mathrm{B}}}{f_2} + {P_{\mathrm{U}}}{f_3}}}, \\
			\max (R_{\mathrm{BS}},R_{\mathrm{UE}}),&\frac{{{P_{\mathrm{B}}}{P_{\mathrm{U}}}{f_1}N + {P_{\mathrm{U}}}{f_3}}}{{{P_{\mathrm{B}}}{P_{\mathrm{U}}}{f_1}N + {P_{\mathrm{B}}}{f_2} + {P_{\mathrm{U}}}{f_3}}} \le \varepsilon \le 1.
		\end{cases}
	\end{align}
	{\noindent} \rule[-0pt]{18.3cm}{0.05em}
\end{figure*}
Based on \eqref{R_D_opt}, \eqref{y_B}, and \eqref{y_U}, the optimal WSR of UL and DL communication is presented in \eqref{y_opt} on the top of next page. The inequality $W_1 < \varepsilon < W_2$ is a necessary but not sufficient condition for the equality $R^\mathrm{opt} = R_{\mathrm{DIS}}^{\mathrm{opt}}$. It is observed that $W_2 - W_1$ increases with $M$, $N$, $P_F$, $P_U$, and $P_B$, which suggests that the practical operating region for the WSR of distributed AIRSs with optimized element allocation that outperforms that of the BS- and user-side AIRS can be extended by increasing the total number of active elements and/or the amplification power budget. It indicates that the distributed AIRS becomes less sensitive to UL/DL weight, significantly enhancing its robustness and flexibility in balancing UL and DL communications. The reasons can be explained as follows. More antennas provide greater spatial degrees of freedom, which enhances spatial diversity. More AIRS elements improve the beamforming gain and allow for more flexible element allocation to satisfy UL and DL requirements. As the maximum amplification power increases, the distributed AIRS can effectively balance signal amplification and noise, while the single AIRS schemes may suffer from inappropriate placement and associated rate loss. Furthermore, higher transmit or amplification power can compensate for the link gain reduction due to element allocation. These results highlight the importance of optimizing AIRS deployment to fully realize the benefits of distributed AIRS-assisted joint UL and DL communications in practical systems.

\section{Multi-User System}
\label{Multi-User System}
Based on the above discussion, we demonstrate the superiority of two distributed AIRSs in the single-user case. In this section, we further investigate the WSR maximization problem in a two distributed AIRSs-aided communication system in the multi-user case. In the following, we proposed two AIRS beamforming setups depending on how the AIRS set its phase shifts over time, namely, user-adaptive and static/constant AIRS beamforming.

\subsection{User-Adaptive AIRS Beamforming}
\label{sec:user-adaptive}
With user-adaptive AIRS beamforming, the AIRSs are allowed to reconfigure their phase-shift patterns/vectors $K$ times and each vector is dedicated to one user. For the minimum achievable rate maximization problem, the system prioritizes the worst-case user. In this case, the problem reduces to an equivalent single-user case, where the analysis and results of the single-user case can be directly applied. In contrast, for the sum-rate maximization problem, the system favors users with better channel conditions to maximize the sum rate, which requires different resource allocation strategies. 

Given the design of AIRS beamforming and BS beamforming similar to the single-user case, the achievable rate of user $k$ in bps/Hz in UL and DL are given by
\begin{align}
	R_{\mathrm{U},k} \!=& \frac{1}{K} {\log _2}\!  \left(  {1 \!+\! \frac{{{P_{\mathrm{U}}}{P_{\mathrm{F}}}Mh_{{\mathrm{U}},k}^2h_1^2 N_\mathrm{U}}}{{{P_{\mathrm{F}}}h_1^2\sigma _{\mathrm{F}}^2 \!+\! {P_{\mathrm{U}}}h_{{\mathrm{U}},k}^2\sigma _0^2 \!+\! \sigma _{\mathrm{F}}^2\sigma _0^2}}}  \right), \\
	R_{\mathrm{D},k} \!=& \frac{1}{K} {\log _2}\!  \left(  {1 \!+\! \frac{{{P_{\mathrm{B}}}{P_{\mathrm{F}}}Mh_{{\mathrm{D}},k}^2h_2^2 N_\mathrm{D}}}{{{P_{\mathrm{F}}}h_{{\mathrm{D}},k}^2\sigma _{\mathrm{F}}^2 \!+\! M{P_{\mathrm{B}}}h_2^2\sigma _0^2 \!+\! \sigma _{\mathrm{F}}^2\sigma _0^2}}}  \right),
\end{align}
where $|h_{{\mathrm{U}},k}|^2$ and $|h_{{\mathrm{D}},k}|^2$ denote the channel gain of the channel from user $k$ to BS-side AIRS, and form user-side AIRS to user $k$, respectively. 
Accordingly, the WSR maximization problem is formulated as
\begin{subequations}
	\label{pro:user-adaptive-N}
	\begin{align}
			\mathop {\max }\limits_{{N_{{\mathrm{U}}}},{N_{{\mathrm{D}}}}} \; &\sum \nolimits_{k = 1}^{K} \left((1-\varepsilon ) R_{\mathrm{U},k} + \varepsilon R_{\mathrm{D},k}\right)
			\\
			\mathrm{s.t.} \;\;\;\;
			&{N_{{\mathrm{U}}}} + {N_{{\mathrm{D}}}} \le N,{N_{{\mathrm{U}}}} \in \mathbb{N}^+,{N_{{\mathrm{D}}}} \in \mathbb{N}^+. \label{N_user-adaptive}
		\end{align}
\end{subequations}
For problem \eqref{pro:user-adaptive-N}, constraint \eqref{N_user-adaptive} is the total number of active elements constraint. Although deriving a solution to problem \eqref{pro:user-adaptive-N} in closed form is challenging, the optimal solution can be efficiently determined through a one-dimensional search over ${N_{{\mathrm{D}}}} \in \left[1,N-1\right]$. In the high SNR case for all users, it can be readily verified that the element allocation under the single-user case is still applicable, i.e., $x^{\mathrm{near-opt}}_\mathrm{D} = \varepsilon N$. 

\subsection{Static/Constant AIRS Beamforming}
In this case, the two AIRSs allowed to reconfigure its beamforming for each user and share the same phase-shift, i.e., ${{\bm{\Phi }}_{{\mathrm{U},k}}} = {{\bm{\Phi }}_{{\mathrm{D},k}}} = {{\bm{\Phi }}} =  \operatorname{\operatorname{diag}} ( {e^{j{\phi _1}}}, \ldots ,{e^{j{\phi _{{N}}}}} )$,$\forall k \in \mathcal{K}$. Moreover, we assume that all reflecting elements at the AIRS share the common amplification factor for each user in UL (DL) communication, i.e., $\alpha _{{\mathrm{U},k}} = \alpha _{{\mathrm{U}}} (\alpha _{{\mathrm{D},k}} = \alpha _{{\mathrm{D}}})$. Accordingly, the achievable rate for UL and DL associated with user $k$ in bps/Hz are given by 
\begin{align}
	R_{\mathrm{U},k} \!\!=& \frac{1}{K}{\log _2} \!\! \left( \!\!1 \!\!+\!\! \frac{{{p_k}{\left| {{\bm{u}}_k^H{\bm{G}}_{{\mathrm{U}}}^H{\alpha _{{\mathrm{U}}}}{{\bm{\Phi }}}{\bm{h}}_{\mathrm{U},k}} \right|^2}}}{{{\bm{u}}_k^H (\! {\alpha _{{\mathrm{U}}}^2\sigma _{\mathrm{F}}^2{\bm{G}}_{{\mathrm{U}}}^H{{\bm{\Phi }}}{\bm{\Phi }}^H{{\bm{G}}_{{\mathrm{U}}}} \!+\! \sigma _0^2{{\bm{I}}_M}} \!) {{\bm{u}}_k}}}\!\! \right) \!,\\
	R_{\mathrm{D},k} \!\!=& \frac{1}{K}{\log _2} \left( {1 + \frac{{{| {{\bm{h}}_{\mathrm{D},k}^H{\alpha _{{\mathrm{D}}}}{{\bm{\Phi }}}{{\bm{G}}_{{\mathrm{D}}}}{{\bm{w}}_k}} |^2}}}{{\alpha _{{\mathrm{D}}}^2\sigma _{\mathrm{F}}^2{\| {{\bm{h}}_{\mathrm{D},k}^H{{\bm{\Phi }}}} \|^2} + \sigma _0^2}}} \right).
\end{align}

We aim to maximize the system’s WSR of the system by jointly optimizing each user’s transmit power, the transmit/receive beamformer of the BS, as well as the AIRS beamforming. The corresponding optimization problem is formulated as
\begin{subequations}
	\label{pro:tdma}
	\begin{align}
		&\!\!\!\!\!\!\!\!\!\!\!\!\!\!\!\!\!\!\!\! \mathop {\max }\limits_{
			{\scriptstyle \{ {{p_k}} \},\{ {{{\bm{w}}_k}} \},{\bm{\Phi }}, \hfill\atop
				\scriptstyle \{{{{\bm{u}}_k}} \},{\alpha _{{\mathrm{U}}}},{\alpha _{{\mathrm{D}}}}\hfill}}
		\sum \nolimits_{k = 1}^{K} \left((1-\varepsilon ) R_{\mathrm{U},k} + \varepsilon R_{\mathrm{D},k}\right)\\
		\mathrm{s.t.} \;\;
		& {{{\left\| {{{\bm{w}}_k}} \right\|}^2}}  \le {P_{\mathrm{B}}},\forall k \in \mathcal{K}, \\
		& {\left\| {{{\bm{u}}_k}} \right\|^2} = 1,\forall k \in \mathcal{K}, \\
		& 0 \le {p_k} \le {P_{\mathrm{U}}},\forall k \in \mathcal{K}, \\
		& \left| {{{\left[ {\bm{\Phi }} \right]}_{n,n}}} \right| = 1,\forall n \in \mathcal{N}, \label{phi} \\
		& \alpha _{{\mathrm{U}}}^2 ( {{{p_k}{\left\| {{{\bm{\Phi }}}{\bm{h}}_{\mathrm{U},k}} \right\|^2}}  + \sigma _{\mathrm{F}}^2\left\| {{{\bm{\Phi }}}} \right\|_F^2} )  \le  {P_{\mathrm{F}}},\! \forall k \in \mathcal{K}, \label{a_ul} \\
		& \alpha _{{\mathrm{D}}}^2 ( {{{\left\| {{{\bm{\Phi }}}{{\bm{G}}_{{\mathrm{D}}}}{{\bm{w}}_k}} \right\|^2}} + \sigma _{\mathrm{F}}^2\left\| {{{\bm{\Phi }}}} \right\|_F^2} ) \le {P_{\mathrm{F}}},\forall k \in \mathcal{K} \label{a_dl}.
	\end{align}
\end{subequations}
For problem \eqref{pro:tdma}, constraint \eqref{phi} is the unit-modulus phase-shift constraint. Constraint \eqref{a_ul} and \eqref{a_dl} ensure that the amplification power of the AIRS for UL and DL communication should not exceed the power budget. Problem \eqref{pro:tdma} is non-convex because the optimization variables are intricately coupled with each other. Different from user-adaptive beamforming, the amplification power constraint at the AIRS under static beamforming, i.e., constraint \eqref{a_ul}, introduces additional complexity. In this case, the amplification factor at the AIRS is coupled with the transmit powers of all users, which implies that increasing the transmit power of one user may reduce the amplification factor, thereby negatively impacting the signal enhancement for all users and potentially degrade the overall system performance. Consequently, it is unclear whether the maximum transmit power should be allocated for each user to maximize the sum rate with static beamforming, which motivates the following proposition.
\begin{Proposition}
	\label{pro:5}
	At the optimal solution to problem \eqref{pro:tdma}, the optimal transmit power of each user is $p_k = P_\mathrm{U}, \forall k \in \mathcal{K}$.
\end{Proposition}
{\it{Proof:}} Please refer to Appendix A.
~$\hfill\blacksquare$

Exploiting Proposition \eqref{pro:5}, the achievable rate for UL associated with user $k$ in bps/Hz can be rewritten as 
\begin{align}
	R_{\mathrm{U},k} \!\!=\!\! \frac{1}{K}{\log _2} \!\! \left( \!\!1 \!\!+\!\! \frac{{{P_\mathrm{U}}{\left| {{\bm{u}}_k^H{\bm{G}}_{{\mathrm{U}}}^H{\alpha _{{\mathrm{U}}}}{{\bm{\Phi }}}{\bm{h}}_{\mathrm{U},k}} \right|^2}}}{{{\bm{u}}_k^H (\! {\alpha _{{\mathrm{U}}}^2\sigma _{\mathrm{F}}^2{\bm{G}}_{{\mathrm{U}}}^H{{\bm{\Phi }}}{\bm{\Phi }}^H{{\bm{G}}_{{\mathrm{U}}}} \!\!+\! \sigma _0^2{{\bm{I}}_M}} \!) {{\bm{u}}_k}}}\!\! \right).
\end{align}
Thus, problem \eqref{pro:tdma} can be simplified to
\begin{subequations}
	\label{pro:new}
	\begin{align}
		&\!\!\!\!\!\!\!\!\!\!\!\!\!\!\!\!\! \mathop {\max }\limits_{
			{\scriptstyle \{ {{{\bm{w}}_k}} \},\{{{\bm{u}}_k} \}, \hfill\atop
				\scriptstyle {\bm{\Phi }},{\alpha _{{\mathrm{U}}}},{\alpha _{{\mathrm{D}}}}\hfill}}
		\sum \nolimits_{k = 1}^{K} \left((1-\varepsilon ) R_{\mathrm{U},k} + \varepsilon R_{\mathrm{D},k}\right)\\
		\mathrm{s.t.} \;\;
		& {{{\left\| {{{\bm{w}}_k}} \right\|}^2}}  \le {P_{\mathrm{B}}},\forall k \in \mathcal{K}, \\
		& {\left\| {{{\bm{u}}_k}} \right\|^2} = 1,\forall k \in \mathcal{K}, \\
		& \left| {{{\left[ {\bm{\Phi }} \right]}_{n,n}}} \right| = 1,\forall n \in \mathcal{N},  \\
		& \alpha _{{\mathrm{U}}}^2 ( {{{P_\mathrm{U}}{\left\| {{{\bm{\Phi }}}{\bm{h}}_{\mathrm{U},k}} \right\|^2}}  + \sigma _{\mathrm{F}}^2\left\| {{{\bm{\Phi }}}} \right\|_F^2} )  \le  {P_{\mathrm{F}}},\! \forall k \in \mathcal{K},  \\
		& \alpha _{{\mathrm{D}}}^2 ( {{{\left\| {{{\bm{\Phi }}}{{\bm{G}}_{{\mathrm{D}}}}{{\bm{w}}_k}} \right\|^2}} + \sigma _{\mathrm{F}}^2\left\| {{{\bm{\Phi }}}} \right\|_F^2} ) \le {P_{\mathrm{F}}},\forall k \in \mathcal{K} .
	\end{align}
\end{subequations}
Despite the non-convexity of problem \eqref{pro:new}, it can be decomposed into five sub-problems and solved via an AO-based algorithm that iteratively optimizes the user, BS, and AIRS beamforming.

\subsubsection{Receive Beamforming Optimization} For any given ${\left\{ \bm{w}_k \right\}}$, ${{\bm{\Phi }}}$, ${\alpha _{{\mathrm{U}}}}$, and ${\alpha _{{\mathrm{D}}}}$, the receive beamforming is designed to maximize the WSR by solving the following problem
\begin{subequations}
	\label{pro:u}
	\begin{align}
		\mathop {\max }\limits_{\left\{ {{{\bm{u}}_k}} \right\}} & \;\;\;
		\sum\nolimits_{k = 1}^K {R_{\mathrm{U},k}} \\
		\mathrm{s.t.}& \;\;\; {\left\| {{{\bm{u}}_k}} \right\|^2} = 1,\forall k \in \mathcal{K}.
	\end{align}
\end{subequations}
Under the LoS channel, the optimal solution to problem \eqref{pro:u} is achieved by the MRC beamforming. For each user $k$, the receive beamforming vector is given by
\begin{align}
	\label{u_opt}
	{{\bm{u}}_k^\mathrm{opt}} \!\!\!=\!\! \frac{{{\bm{a }}_{\mathrm{t}} ( {\theta _{{\mathrm{BI,UL}}}^{\mathrm{D}},\! \nu _{{\mathrm{BI,UL}}}^{\mathrm{D}},\! M} \!){{\bm{a }}_{\mathrm{r}}}^H ( {\theta _{{\mathrm{BI,UL}}}^{\mathrm{A}},\! \nu _{{\mathrm{BI,UL}}}^{\mathrm{A}},\! {N_{{\mathrm{U}}}}} \!)} {\bf{\Phi }}{{\bf{h}}_{{\rm{U}},k}}}{{\left\| {{\bm{a }}_{\mathrm{t}} ( {\theta _{{\mathrm{BI,UL}}}^{\mathrm{D}},\! \nu _{{\mathrm{BI,UL}}}^{\mathrm{D}},\! M} \!){{\bm{a }}_{\mathrm{r}}}^H ( {\theta _{{\mathrm{BI,UL}}}^{\mathrm{A}},\! \nu _{{\mathrm{BI,UL}}}^{\mathrm{A}},\! {N_{{\mathrm{U}}}}} \!)} {\bf{\Phi }}{{\bf{h}}_{{\rm{U}},k}} \right\|}}.
\end{align}

\subsubsection{Transmit Beamforming Optimization} For any given ${\left\{ {{{\bm{u}}_k}} \right\}}$, ${{\bm{\Phi }}}$, ${\alpha _{{\mathrm{U}}}}$, and ${\alpha _{{\mathrm{D}}}}$, the optimization problem w.r.t. transmit beamforming can be written as
\begin{subequations}
	\label{pro:w}
	\begin{align}
		\mathop {\max }\limits_{\left\{ {{{\bm{w}}_k}} \right\}} \; 
		& \sum\nolimits_{k = 1}^K {R_{\mathrm{D},k}} \\
		\mathrm{s.t.} \;\; 
		& {{{\left\| {{{\bm{w}}_k}} \right\|}^2}}  \le {P_{\mathrm{B}}},\forall k \in \mathcal{K}, \\
		&\alpha _{{\mathrm{D}}}^2 ( {{{{\left\| {{{\bm{\Phi }}}{{\bm{G}}_{{\mathrm{D}}}}{{\bm{w}}_k}} \right\|}^2}} \!+\! \sigma _{\mathrm{F}}^2\left\| {{{\bm{\Phi }}}} \right\|_F^2} ) \le {P_{\mathrm{F}}},\forall k \in \mathcal{K}.
	\end{align}
\end{subequations}
Let ${\bm{Q}} = {\bm{G}}_{{\mathrm{D}}}^H{{\bm{\Phi }}^H}{{\bm{h}}_k^\mathrm{D}}({\bm{h}}_k^\mathrm{D})^H{\bm{\Phi }}{{\bm{G}}_{{\mathrm{D}}}}$ and $\tilde {\bm{Q}} = \alpha _{{\mathrm{D}}}^2 {\bm{G}}_{{\mathrm{D}}}^H{{\bm{\Phi }}^H} {\bm{\Phi }}{{\bm{G}}_{{\mathrm{D}}}}$, problem \eqref{pro:w} can be reformulated in an equivalent form as 
\begin{subequations}
	\label{pro:W}
	\begin{align}
		\mathop {\max }\limits_{\left\{ {{{\bm{w}}_k}} \right\}} \; 
		& \bm{w}_k^H \bm{Q} \bm{w}_k \\
		\mathrm{s.t.} \;\; 
		& {{{\left\| {{{\bm{w}}_k}} \right\|}^2}}  \le {P_{\mathrm{B}}},\forall k \in \mathcal{K}, \\
		& \bm{w}_k^H \tilde {\bm{Q}} \bm{w}_k \le {P_{\mathrm{F}}} - \alpha _{{\mathrm{D}}}^2\sigma _{\mathrm{F}}^2\left\| {\bm{\Phi }} \right\|_F^2, \forall k \in \mathcal{K}.
	\end{align}
\end{subequations}
Problem \eqref{pro:W} is still a non-convex optimization problem. Define $\bm{W}_k = \bm{w}_k\bm{w}_k^H, \forall k$, which satisfy $\bm{W}_k \succeq 0$ and $\operatorname{rank}(\bm{W}_k) = 1$. After dropping the rank-one constraint, the SDR of problem \eqref{pro:W} is given by
\begin{subequations}
	\label{pro:W_1}
	\begin{align}
		\mathop {\max }\limits_{\left\{ {{{\bm{W}}_k}} \right\}} \; 
		& \operatorname{tr}(\bm{Q}\bm{W}_k)  \\
		\mathrm{s.t.} \;\; 
		& \operatorname{tr}(\bm{W}_k)  \le {P_{\mathrm{B}}},\forall k \in \mathcal{K}, \\
		& \operatorname{tr}(\tilde {\bm{Q}} \bm{W}_k) \le {P_{\mathrm{F}}} - \alpha _{{\mathrm{D}}}^2\sigma _{\mathrm{F}}^2\left\| {\bm{\Phi }} \right\|_F^2, \forall k \in \mathcal{K}, \\
		& \bm{W}_k \succeq 0.
	\end{align}
\end{subequations}
Problem \eqref{pro:W_1} is standard semidefinite program (SDP), which can be solved by CVX. To recover the rank-one solution, we can obtain $\bm{w}_k$ through Cholesky decomposition.

\subsubsection{IRS Amplification Factor Optimization for Uplink} For any given ${\left\{ {{{\bm{w}}_k}} \right\}}$, ${\left\{ {{{\bm{u}}_k}} \right\}}$, ${{\bm{\Phi }}}$, and ${\alpha _{{\mathrm{D}}}}$, the AIRS amplification factor optimization problem for UL is formulated as
\begin{subequations}
	\label{pro:a_ul}
	\begin{align}
		\mathop {\max }\limits_{{\alpha _{{\mathrm{U}}}}} \;\;&\sum\nolimits_{k = 1}^K {R_{\mathrm{U},k}}\\
		\mathrm{s.t.} \;\;
		& \alpha _{{\mathrm{U}}}^2 ( {{{P_\mathrm{U}}{{\left\| {{{\bm{\Phi }}}{\bm{h}}_{\mathrm{U},k}} \right\|}^2}}  \!+\! \sigma _{\mathrm{F}}^2\left\| {{{\bm{\Phi }}}} \right\|_F^2} ) \le {P_{\mathrm{F}}},\forall k \in \mathcal{K} \label{con:a_ul}.
	\end{align}
\end{subequations}
For problem \eqref{pro:a_ul}, constraint \eqref{con:a_ul} is strictly satisfied with equality at the optimal solution because the objective value can always be improved by increasing $\alpha _{{\mathrm{U}}}$ until constraint \eqref{con:a_ul} becomes active. Therefore, the optimal solution to problem \eqref{pro:a_ul} is expressed as
\begin{align}
	\label{a_ul_opt}
	{\alpha _{{\mathrm{U}}}^\mathrm{opt}} = \min \left\{{\sqrt {\frac{{{P_{\mathrm{F}}}}}{{{{P_\mathrm{U}}{\| {{\bm{\Phi h}}_{\mathrm{U},k}} \|^2} + \sigma _{\mathrm{F}}^2 N } }}} }, \forall k \in \mathcal{K} \right\}.
\end{align}

\subsubsection{IRS Amplification Factor Optimization for Downlink} For any given ${\left\{ {{{\bm{w}}_k}} \right\}}$, ${\left\{ {{{\bm{u}}_k}} \right\}}$, ${{\bm{\Phi }}}$, and ${\alpha _{{\mathrm{U}}}}$, the AIRS amplification factor for DL is optimized by solving the following problem
\begin{subequations}
	\label{pro:a_dl}
	\begin{align}
		\mathop {\max }\limits_{{\alpha _{{\mathrm{D}}}}} \;\;\; &\sum\nolimits_{k = 1}^K {R_{\mathrm{D},k}} \\
		\mathrm{s.t.} \;\;\;\;
		&\alpha _{{\mathrm{D}}}^2 ( {{{{\left\| {{{\bm{\Phi }}}{{\bm{G}}_{{\mathrm{D}}}}{{\bm{w}}_k}} \right\|}^2}} \!+\! \sigma _{\mathrm{F}}^2\left\| {{{\bm{\Phi }}}} \right\|_F^2} ) \le {P_{\mathrm{F}}},\forall k \in \mathcal{K} \label{con:a_dl}.
	\end{align}
\end{subequations}
For problem \eqref{pro:a_dl}, constraint \eqref{con:a_dl} is strictly met with equality at the optimal solution since increasing $\alpha _{{\mathrm{D}}}$ improves the objective value until constraint \eqref{con:a_dl} is active. Therefore, the optimal solution to problem \eqref{pro:a_dl} is expressed as
\begin{align}
	\label{a_dl_opt}
	{\alpha _{{\mathrm{D}}}^\mathrm{opt}} = \min \left\{{\sqrt {\frac{{{P_{\mathrm{F}}}}}{{{{\left\| {{\bm{\Phi }}{{\bm{G}}_{{\mathrm{D}}}}{{\bm{w}}_k}} \right\|}^2} + \sigma _{\mathrm{F}}^2N}}} }, \forall k \in \mathcal{K}\right\}.
\end{align}

\subsubsection{IRS Phase-Shift Optimization} For any given ${\left\{ {{{\bm{w}}_k}} \right\}}$, ${\left\{ {{{\bm{u}}_k}} \right\}}$, ${\alpha _{{\mathrm{U}}}}$, and ${\alpha _{{\mathrm{D}}}}$, the phase shifts of the AIRSs are designed to maximized the WSR by solving the following problem
\begin{subequations}
	\label{pro:phi}
	\begin{align}
		\mathop {\max }\limits_{\bm{\Phi }} \; &( {1 - \varepsilon } )\sum\nolimits_{k = 1}^K {R_{\mathrm{U},k}}  + \varepsilon \sum\nolimits_{k = 1}^K {R_{\mathrm{D},k}}  \\
		\mathrm{s.t.} \;\;
		& | {{{\left[ {\bm{\Phi }} \right]}_{n,n}}} | = 1,\forall n \in \mathcal{N}. \label{phi_unit}
	\end{align}
\end{subequations}
By applying the Lagrangian Dual Transform \cite{QT}, the objective function of problem \eqref{pro:phi} is equivalent to
\begin{align}
	\label{f1}
	&{f_1}( {{\bm{\Phi }},{\bm{\bar \mu }},{\bm{\tilde \mu }}} ) 
	= \frac{1 - \varepsilon }{K \log 2} \sum\nolimits_{k = 1}^K ( \log ( {1 + {{\bar \mu }_k}} ) - {{\bar \mu }_k} \nonumber \\
	&+ \frac{{( {1 + {{\bar \mu }_k}} ){P_\mathrm{U}}{| {{\bm{u}}_k^H{\bm{G}}_{{\mathrm{U}}}^H{\alpha _{{\mathrm{U}}}}{\bm{\Phi h}}_{\mathrm{U},k}} |^2}}}{{{P_\mathrm{U}}{{| {{\bm{u}}_k^H{\bm{G}}_{{\mathrm{U}}}^H{\alpha _{{\mathrm{U}}}}{\bm{\Phi h}}_{\mathrm{U},k}} |}^2} + \alpha _{{\mathrm{U}}}^2\sigma _{\mathrm{F}}^2{{\| {{\bm{u}}_k^H{\bm{G}}_{{\mathrm{U}}}^H{\bm{\Phi }}} \|}^2} + \sigma _0^2}} )  \nonumber\\
	&+ \frac{\varepsilon}{K \log 2} \sum\nolimits_{k = 1}^K ( \log ( {1 + {{\tilde \mu }_k}} ) - {{\tilde \mu }_k} \nonumber\\
	&+ \frac{{( {1 + {{\tilde \mu }_k}} ){| {{\bm{h}}_{\mathrm{D},k}^H{\alpha _{{\mathrm{D}}}}{\bm{\Phi }}{{\bm{G}}_{{\mathrm{D}}}}{{\bm{w}}_k}} |^2}}}{{{| {{\bm{h}}_{\mathrm{D},k}^H{\alpha _{{\mathrm{D}}}}{\bm{\Phi }}{{\bm{G}}_{{\mathrm{D}}}}{{\bm{w}}_k}} |^2} + \alpha _{{\mathrm{D}}}^2\sigma _{\mathrm{F}}^2{ \| {{\bm{h}}_{\mathrm{D},k}^H{\bm{\Phi }}} \|^2} + \sigma _0^2}} ),
\end{align}
when the auxiliary vectors ${\bm{\bar \mu}} \triangleq \{ {{\bar \mu }_1},{{\bar \mu }_2}, \cdots ,{{\bar \mu }_K} \}$ and ${\bm{\tilde \mu }} \triangleq \{ {{\tilde \mu }_1},{{\tilde \mu }_2}, \cdots ,{{\tilde \mu }_K} \}$ has the optimal solution as
\begin{align}
	\bar \mu _k^{{\mathrm{opt}}} =& \frac{{{P_\mathrm{U}}{| {{\bm{u}}_k^H{\bm{G}}_{{\mathrm{U}}}^H{\alpha _{{\mathrm{U}}}}{\bm{\Phi h}}_{\mathrm{U},k}} |^2}}}{{ \alpha _{{\mathrm{U}}}^2\sigma _{\mathrm{F}}^2{\| {{\bm{u}}_k^H{\bm{G}}_{{\mathrm{U}}}^H{\bm{\Phi }}} \|^2} + \sigma _0^2}}, \label{mu_bar_opt}\\
	\tilde \mu _k^{{\mathrm{opt}}} =& \frac{{{| {{\bm{h}}_{\mathrm{D},k}^H{\alpha _{{\mathrm{D}}}}{\bm{\Phi }}{{\bm{G}}_{{\mathrm{D}}}}{{\bm{w}}_k}} |^2}}}{{\alpha _{{\mathrm{D}}}^2\sigma _{\mathrm{F}}^2{\| {{\bm{h}}_{\mathrm{D},k}^H{\bm{\Phi }}} \|^2} + \sigma _0^2}} \label{mu_tilde_opt}.
\end{align}
Since two terms of \eqref{f1} is a sum of multiple
fractions, the corresponding optimization problem is still non-convex and challenging to be solved. To address this, we adopt Quadratic Transform technique \cite{QT} and rewritten \eqref{f1} as
\begin{align}
	&{f_2}( {{\bm{\Phi }},{\bm{\bar \mu }},{\bm{\tilde \mu }},{\bm{\bar \eta }},{\bm{\tilde \eta }}} ) 
	= \frac{1 - \varepsilon }{K \log 2} \sum\nolimits_{k = 1}^K ( \log ( {1 + {{\bar \mu }_k}} ) - {{\bar \mu }_k} \nonumber\\
	&+ 2\sqrt {1 + {{\bar \mu }_k}} \Re \{ \bar \eta _k^* \sqrt{P_\mathrm{U}} {\bm{u}}_k^H{\bm{G}}_{{\mathrm{U}}}^H{\alpha _{{\mathrm{U}}}}{\bm{\Phi h}}_{\mathrm{U},k} \} \nonumber\\
	&- {| {{{\bar \eta }_k}} |^2} ( {\alpha _{{\mathrm{U}}}^2}{P_\mathrm{U}}{| {{\bm{u}}_k^H{\bm{G}}_{{\mathrm{U}}}^H{\bm{\Phi h}}_{\mathrm{U},k}} |^2} \!+\! \alpha _{{\mathrm{U}}}^2\sigma _{\mathrm{F}}^2{\| {{\bm{u}}_k^H{\bm{G}}_{{\mathrm{U}}}^H{\bm{\Phi }}} \|^2} \!+\! \sigma _0^2 ) )  \nonumber\\
	&+ \frac{\varepsilon}{K \log 2} \sum\nolimits_{k = 1}^K ( \log ( {1 + {{\tilde \mu }_k}} ) - {{\tilde \mu }_k} \nonumber\\
	&+ 2\sqrt {1 + {{\tilde \mu }_k}} \Re \{ {\tilde \eta _k^*{\bm{h}}_{\mathrm{D},k}^H{\alpha _{{\mathrm{D}}}}{\bm{\Phi }}{{\bm{G}}_{{\mathrm{D}}}}{{\bm{w}}_k}} \} \nonumber\\
	&- {| {{{\tilde \eta }_k}} |^2} ( {{{\alpha _{{\mathrm{D}}}^2}| {{\bm{h}}_{\mathrm{D},k}^H{\bm{\Phi }}{{\bm{G}}_{{\mathrm{D}}}}{{\bm{w}}_k}} |^2} \!+\! \alpha _{{\mathrm{D}}}^2\sigma _{\mathrm{F}}^2{\| {{\bm{h}}_{\mathrm{D},k}^H{\bm{\Phi }}} \|^2} \!+\! \sigma _0^2} ) ), 
\end{align}
when the auxiliary vectors ${\bm{\bar \eta}} \triangleq \{ {{\bar \eta }_1},{{\bar \eta }_2}, \cdots ,{{\bar \eta }_K} \}$ and ${\bm{\tilde \eta }} \triangleq \{ {{\tilde \eta }_1},{{\tilde \eta }_2}, \cdots ,{{\tilde \eta }_K} \}$ has the optimal solution as
\begin{align}
	\bar \eta _k^{{\mathrm{opt}}} =& \frac{{\sqrt {1 + {{\bar \mu }_k}} \sqrt{P_\mathrm{U}} {\bm{u}}_k^H{\bm{G}}_{{\mathrm{U}}}^H{\alpha _{{\mathrm{U}}}}{\bm{\Phi h}}_{\mathrm{U},k}}}{{{P_\mathrm{U}}{| {{\bm{u}}_k^H{\bm{G}}_{{\mathrm{U}}}^H{\alpha _{{\mathrm{U}}}}{\bm{\Phi h}}_{\mathrm{U},k}} |^2} + \alpha _{{\mathrm{U}}}^2\sigma _{\mathrm{F}}^2{\| {{\bm{u}}_k^H{\bm{G}}_{{\mathrm{U}}}^H{\bm{\Phi }}} \|^2} + \sigma _0^2}}, \label{eta_bar_opt}\\
	\tilde \eta _k^{{\mathrm{opt}}} =& \frac{{\sqrt {1 + {{\tilde \mu }_k}} {\bm{h}}_{\mathrm{D},k}^H{\alpha _{{\mathrm{D}}}}{\bm{\Phi }}{{\bm{G}}_{{\mathrm{D}}}}{{\bm{w}}_k}}}{{{| {{\bm{h}}_{\mathrm{D},k}^H{\alpha _{{\mathrm{D}}}}{\bm{\Phi }}{{\bm{G}}_{{\mathrm{D}}}}{{\bm{w}}_k}} |^2} + \alpha _{{\mathrm{D}}}^2\sigma _{\mathrm{F}}^2{\| {{\bm{h}}_{\mathrm{D},k}^H{\bm{\Phi }}} \|^2} + \sigma _0^2}} \label{eta_tilde_opt}.
\end{align}
Let $\bm{v} = \left[v_1, \cdots, v_N\right]^H$ where $v_n = e^{j \phi_n}, \forall n \in \mathcal{N}$. Given ${\bm{\bar \mu }}$, ${\bm{\tilde \mu }}$, ${\bm{\bar \eta }}$, ${\bm{\tilde \eta }}$, problem \eqref{pro:phi} can be reformulated as
\begin{subequations}
	\label{pro:phi_3}
	\begin{align}
		\mathop {\max }\limits_{\bm{v}} &\;\; - {{\bm{v}}^H}{\bm{Av}} + 2 \Re \left\{ {{{\bm{v}}^H}{\bm{b}}} \right\}  \\
		\mathrm{s.t.}& \;\; \left|v_n\right| = 1, \forall n \in \mathcal{N},
	\end{align}
\end{subequations}
with ${\bm{b}} = \sum\nolimits_{k = 1}^K (\varepsilon \sqrt {1 + {{\tilde \mu }_k}} {\alpha _{{\mathrm{D}}}}\tilde \eta _k^*\operatorname{diag}( {{ {\bm{h}}_{\mathrm{D},k}^H}} ){{\bm{G}}_{{\mathrm{D}}}}{{\bm{w}}_k} + ( {1 - \varepsilon } )\sqrt {1 + {{\bar \mu }_k}} \sqrt{P_\mathrm{U}} \bar \eta _k^* {\alpha _{{\mathrm{U}}}} \operatorname{diag}( {{\bm{u}}_k^H{\bm{G}}_{{\mathrm{U}}}^H} ){\bm{h}}_{\mathrm{U},k} )$ and
\begin{align}
	{\bm{A}} &= ( {1 - \varepsilon } )\sum\nolimits_{k = 1}^K ({P_\mathrm{U}}{\left| {{{\bar \eta }_k}} \right|^2}\alpha _{{\mathrm{U}}}^2 \operatorname{diag}( {{\bm{u}}_k^H{\bm{G}}_{{\mathrm{U}}}^H} ){\bm{h}}_{\mathrm{U},k}{ {\bm{h}}_{\mathrm{U},k}^H}  \nonumber \\ 
	&\times \operatorname{diag}( {{{\bm{G}}_{{\mathrm{U}}}}{{\bm{u}}_k}} ) + {{\left| {{{\bar \eta }_k}} \right|}^2}\alpha _{{\mathrm{U}}}^2\sigma _{\mathrm{F}}^2\operatorname{diag}( {{\bm{u}}_k^H{\bm{G}}_{{\mathrm{U}}}^H} ) \operatorname{diag}( {{{\bm{G}}_{{\mathrm{U}}}}{{\bm{u}}_k}} ) )\nonumber\\
	&+ \varepsilon \sum\nolimits_{k = 1}^K ({{\left| {{{\tilde \eta }_k}} \right|}^2}\alpha _{{\mathrm{D}}}^2 \operatorname{diag}( { {{\bm{h}}_{\mathrm{D},k}^H}} ) {{\bm{G}}_{{\mathrm{D}}}}{{\bm{w}}_k}{\bm{w}}_k^H{\bm{G}}_{{\mathrm{D}}}^H \operatorname{diag}( {{\bm{h}}_{\mathrm{D},k}})  \nonumber\\
	&+{{\left| {{{\tilde \eta }_k}} \right|}^2}\alpha _{{\mathrm{D}}}^2\sigma _{\mathrm{F}}^2 \operatorname{diag}( { {\bm{h}}_{\mathrm{D},k} ^H} )  \operatorname{diag}( {{\bm{h}}_{\mathrm{D},k}} )).
\end{align}
Problem \eqref{pro:phi_3} is a non-convex quadratically constrained quadratic programming problem, and by introducing an auxiliary variable $\tau$, it can be equivalently written as
\begin{subequations}
	\label{pro:phi_4}
	\begin{align}
		\mathop {\max }\limits_{\bar {\bm{v}}} &\;\; - {\bar {\bm{v}}^H}  {\hat {\bm{Q}}} {\bar {\bm{v}}} \\
		\mathrm{s.t.} &\;\; \left|\bar {\bm{v}}_n\right| = 1, \forall n \in \mathcal{N},
	\end{align}
\end{subequations}
with
\begin{align}
	{\hat {\bm{Q}}} = \left[ {\begin{array}{*{20}{c}}
			{\bm{A}}&{ - {\bm{b}}}\\
			{ - {{\bm{b}}^H}}&0
	\end{array}} \right], \bar {\bm{v}} = \left[ {\begin{array}{*{20}{c}}
	{\bm{v}}\\
	\tau
\end{array}} \right].
\end{align}
However, problem \eqref{pro:phi_4} is generally NP-hard. Define $\bm{V} = \bar {\bm{v}} \bar {\bm{v}}^H$, which satisfies $\bm{V} \succeq 0$ and $\operatorname{rank} (\bm{V}) = 1$. Due to the non-convex rank-one constraint, SDR is applied to relax it. As such, problem \eqref{pro:phi_4} is reduced to
\begin{subequations}
	\label{pro:phi_5}
	\begin{align}
		\mathop {\min }\limits_{{\bm{V}}} \;\; &\operatorname{tr}({\hat {\bm{Q}}} \bm{V})\\
		\mathrm{s.t.} \;\;
		&{\bm{V}}_{n,n} = 1, \forall n \in \mathcal{N}, \\
		&\bm{V} \succeq 0.
	\end{align}
\end{subequations}
Problem \eqref{pro:phi_5} is a SDP that can be solved using by existing convex optimization solvers, e.g., CVX. If the solution is not rank-one, the standard Gaussian randomization technique can be applied to generate a high-quality near-optimal solution to problem \eqref{pro:phi_3}, i.e., $\bm{v} = e^{j \arg(\frac{\bar {\bm{v}} {[1:N]}}{\bar {\bm{v}}_{N+1}} )}$, where $\bar {\bm{v}}{[1:N]}$ denotes the first $N$ items of $\bar {\bm{v}}$.

\subsubsection{Overall Algorithm and Computational Complexity Analysis}
\begin{algorithm}[t]
	\label{Algo: 1}
	\SetAlgoLined 
	\caption{Alternating Optimization Algorithm}
	\KwIn{$P_\mathrm{U}$, $P_\mathrm{B}$, $P_\mathrm{F}$, $M$, $N$, $\varepsilon$, $L$, $H$, $D$, $\mathbf{G}_\mathrm{U}$, $\mathbf{G}_\mathrm{D}$, ${\bm{h}}_k^\mathrm{U}$, ${\bm{h}}_k^\mathrm{D}$, ${{\sigma _0^2}}$, ${{\sigma _\mathrm{F}^2}}$.}
	\KwOut{$\{ {{{\bm{w}}_k^\mathrm{opt}}} \}$, ${\bm{\Phi }^\mathrm{opt}}$, $\{{{{\bm{u}}_k^\mathrm{opt}}} \}$, ${\alpha _{{\mathrm{U}}}^\mathrm{opt}}$, ${\alpha _{{\mathrm{D}}}^\mathrm{opt}}$.}
	Initialize ${\bm{\Phi }}$, ${\alpha _{{\mathrm{U}}}}$, ${\alpha _{{\mathrm{D}}}}$.\\
	\Repeat{convergence} 
	{
		Update ${{{\bm{u}}_k}} $ by \eqref{u_opt} and $ {{{\bm{w}}_k}} $ by sloving problem \eqref{pro:W_1} and performing Cholesky decomposition;\\
		Update ${\alpha _{{\mathrm{U}}}}$ by \eqref{a_ul_opt} and ${\alpha _{{\mathrm{D}}}}$ by \eqref{a_dl_opt};\\
		\Repeat{convergence} 
		{
			Update ${\bm{\bar \mu }}$ and ${\bm{\tilde \mu }}$ by \eqref{mu_bar_opt} and \eqref{mu_tilde_opt}; \\
			Update ${\bm{\bar \eta }}$ and ${\bm{\tilde \eta }}$ by \eqref{eta_bar_opt} and \eqref{eta_tilde_opt};\\
			Update ${\bm{\Phi }}$ by sovling problem \eqref{pro:phi_5} and applying Gaussian randomization;
		}
	}
	Return ${{{\bm{w}}_k^\mathrm{opt}}} = {{{\bm{w}}_k}} $, ${\bm{\Phi }^\mathrm{opt}} = {\bm{\Phi }}$, ${{{\bm{u}}_k^\mathrm{opt}}} ={{{\bm{u}}_k}}$, ${\alpha _{{\mathrm{U}}}^\mathrm{opt}} = {\alpha _{{\mathrm{U}}}}$, ${\alpha _{{\mathrm{D}}}^\mathrm{opt}} = {\alpha _{{\mathrm{D}}}}$.
\end{algorithm}
A two-layer AO algorithm is proposed based on the solutions to the sub-problems. The inner-layer AO updates ${\bm{\Phi }}$ by iteratively updating auxiliary variables and solving problem \eqref{pro:phi_5}. In the outer layer, we solve problem \eqref{pro:new} via \eqref{u_opt}, \eqref{a_ul_opt}, \eqref{a_dl_opt}, and solving problems \eqref{pro:w} and \eqref{pro:phi} alternately in an iterative manner until convergence. The main procedures for solving problem \eqref{pro:new} are summarized in Algorithm \ref{Algo: 1}. 

In each outer iteration, the complexity order of \eqref{u_opt}, \eqref{a_ul_opt}, and \eqref{a_dl_opt} is given by $\mathcal{O}\left(KNM^2+KM^{3}\right)$, $\mathcal{O}\left(KN\right)$, and $\mathcal{O}\left(KMN\right)$. For transmit beamforming design, the computational complexity for solving problem \eqref{pro:w} is $\mathcal{O}\left(KM^{3.5}\right)$. In each inner iteration, solving sub-problem with respect to ${\bm{\Phi }}$ requires the complexities $\mathcal{O}\left(N^{3.5}\right)$. The complexity of updating the closed forms ${\bm{\bar \mu }}$, ${\bm{\tilde \mu }}$, ${\bm{\bar \eta }}$, and ${\bm{\tilde \eta }}$ is $\mathcal{O}\left( KMN \right)$. Overall, the total complexity of the Algorithm \ref{Algo: 1} is $\mathcal{O}\left(I_\mathrm{o}\left(KNM^2+KM^{3.5}+I_{\mathrm{i},l}\left(KMN+N^{3.5}\right)\right)\right)$, where $I_\mathrm{o}$ and $I_{\mathrm{i},l}$ are the number of iterations required for convergence of the outer loop and the inner loop in the $l$-th outer loop, respectively.

\section{Simulation Results}
\label{Simulation}
In this section, numerical results are presented to demonstrate the effectiveness of the proposed distributed AIRS architecture. The BS, BS-side AIRS, and user-side AIRS are positioned at (0, 0, 0) meter (m), (0, 0, $H$) m, and (0, $D$, $H$) m, respectively. The reference channel power gain is set as -30 dB. All links are assumed to be free-space LoS channels. Other parameters are set as follows: $P_U = 15$ dBm, $P_B = 20$ dBm, $P_\mathrm{F} = -5$ dBm, $\varepsilon = 0.4$, $D = 200$ m, $H = 10$ m, $M = 4$, and $\sigma_{\mathrm{F}}^2 = \sigma_0^2 = -80$ dBm. To facilitate performance comparison, we consider the single PIRS-aided system, where the PIRS consists of $N$ passive elements and is deployed at (0, 0, 0) m. 

\subsection{Single-User System}
We first study the WSR performance of the distributed AIRSs-aided communication system, considering the special case with one single user located at (0, $D$, 0).

\subsubsection{Performance Comparison for Different Deployment Schemes}
\begin{figure}[t]
	\centering
	\includegraphics[width=0.95\linewidth]{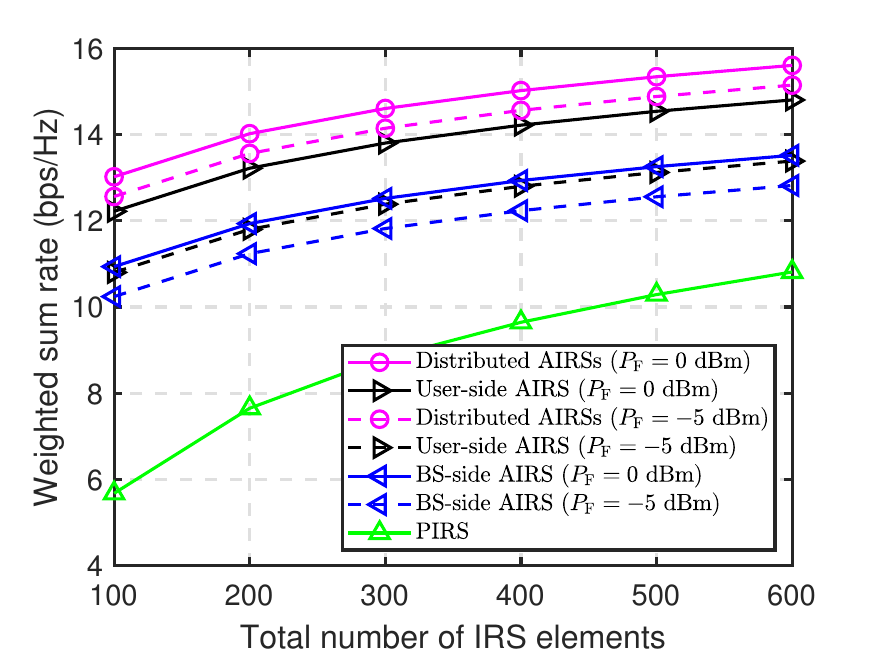}
	\caption{WSR versus total number of IRS elements $N$ under the single-user setup.}
	\label{fig:N_single}
\end{figure}
In Fig. \ref{fig:N_single}, the WSR versus the number of IRS elements $N$ is plotted for both $P_\mathrm{F} = 0$ dBm and $P_\mathrm{F} = -5$ dBm. One can observe that the WSR of all schemes monotonically increases as $N$ increases, since a larger number of passive/active elements enhances IRS beamforming gains, thereby improving received signal power. Moreover, it is observed that all schemes with AIRSs can achieve significant gains over employing PIRS in terms of the WSR thanks to the amplification gain. The performance gap is more significant in the small $N$ regime. In addition, the WSR of AIRS-aided systems with $P_\mathrm{F} = -5$ dBm is lower than that with $P_\mathrm{F} = 0$ dBm. This is due to the reduced amplification capability at each AIRS element caused by the lower value of $P_\mathrm{F}$, which introduces challenges in balancing signal and noise amplification. Compared to the BS/user-side AIRS, the distributed AIRS scheme with an optimized number of active elements achieves superior rate performance, particularly under the amplification power-limited case, i.e., $P_\mathrm{F} = 0$ dBm. The reason is that the distributed AIRSs offer an additional degree of freedom for allocating the number of active elements between UL and DL communications when their allocated times are comparable. This suggests that the distributed AIRS architecture offers greater potential for supporting join UL and DL communications compared to a single AIRS.

\begin{figure}[t]
	\centering
	\includegraphics[width=0.95\linewidth]{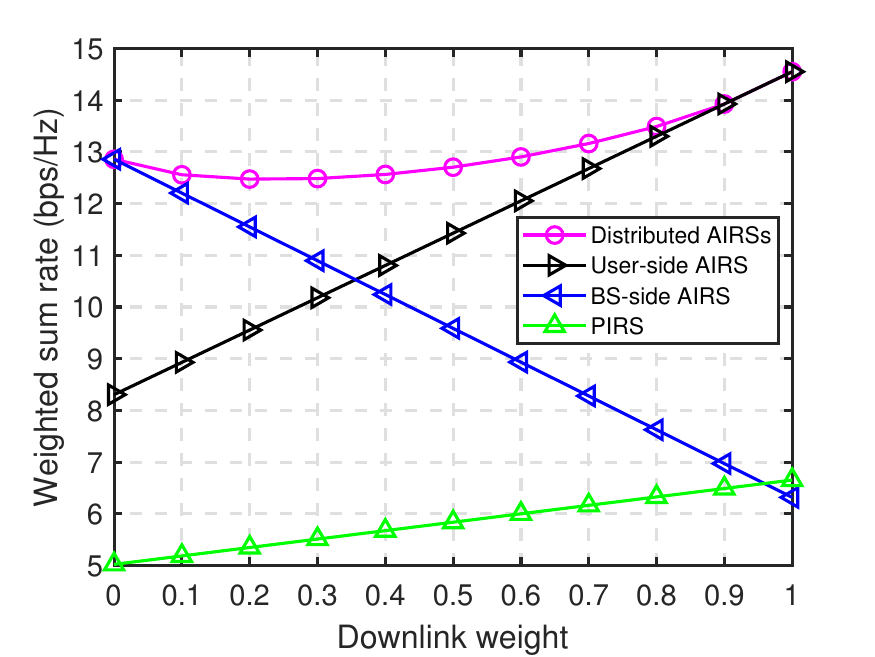}
	\caption{WSR versus DL weight $\varepsilon$ under the single-user setup.}
	\label{fig:t_single}
\end{figure}

In Fig. \ref{fig:t_single}, we plot the WSR versus DL weight when $N = 100$. It is observed that the distributed AIRS outperforms the BS/user-side AIRSs and the PIRS, under different DL weights. This is because the distributed AIRSs are deployed closer to receivers in both UL and DL communications, which maximizes their amplification factors to compensate the first-hop path loss. This deployment efficiently matches the asymmetry of UL and DL channels, thereby further improving rate performance. Moreover, its WSR first decreases and then increases as $\varepsilon$ increases. This is because UL rate loss dominates in the low-$\varepsilon$ regime, while DL beamforming gain outweigh UL penalties when the normalized DL weight exceeds a threshold. The WSR of the distributed AIRS is equal to that of the BS-side AIRS and the user-side AIRS with all the active elements are allocated for UL and DL communications when $\varepsilon = 0$ and $\varepsilon = 1$, respectively. The results highlight the importance of carefully optimizing the AIRS element allocation for the distributed AIRS scheme.

\subsubsection{Performance Evaluation for Distributed AIRS Scheme}
\begin{figure}[t]
	\centering
	\includegraphics[width=0.95\linewidth]{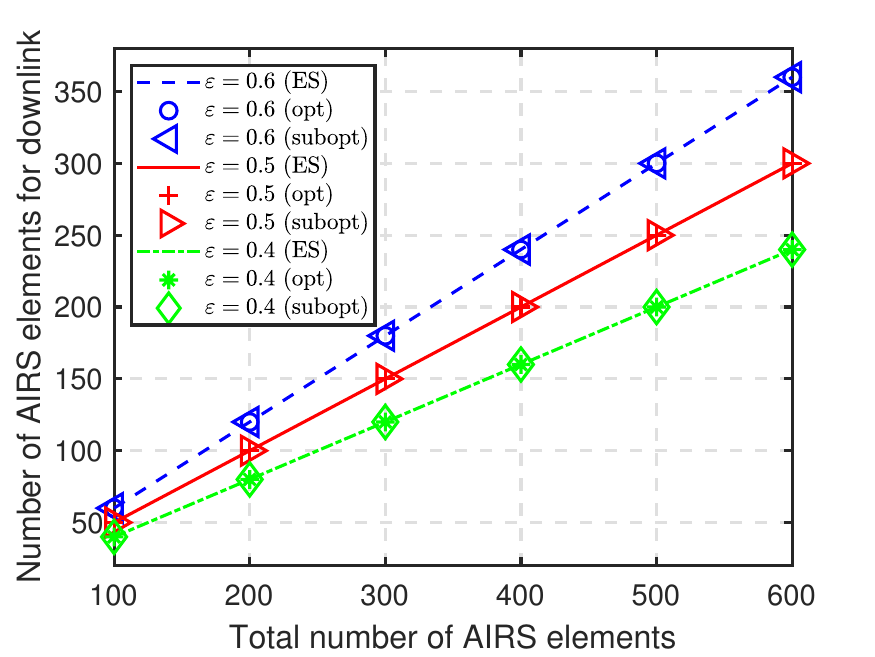}
	\caption{Number of AIRS elements for downlink $N_\mathrm{D}$ versus total number of AIRS elements $N$ under the single-user setup.}
	\label{fig:N_DL}
\end{figure}

In Fig. \ref{fig:N_DL}, we provide performance evaluation for the distributed AIRS scheme by plotting the number of AIRS elements for DL communication versus total number of AIRS elements for $\varepsilon = 0.4$, $\varepsilon = 0.5$, and $\varepsilon = 0.6$. For comparison, we compare the following three cases: 1) \textbf{ES}: exhaustive search is performed for the AIRS element allocation; 2) \textbf{opt}: the element allocation is optimized based on \eqref{x_DL}; 3) \textbf{subopt}: the element allocation is optimized based on \eqref{x_DL_subopt}. As shown in Fig. \ref{fig:N_DL}, it can be observed that the optimized number of AIRS elements for DL increases linearly with the total number of AIRS elements and increases with $\varepsilon$. Moreover, the subopt scheme yields near-optimal allocation compared to both the proposed optimal solution and ES, which demonstrates its effectiveness.

\subsection{Multi-User System}
Next, we study the sum rate of the distributed AIRSs-aided communication system under the multi-user case with $K = 4$. The users are uniformly distributed within a circular area centered at (0, $D$, 0) m with a diameter of 10 m. To facilitate comparison, the user-adpative IRS beamforming is adopted at the user-side AIRS, BS-side AIRS and PIRS schemes, which share the same total number of IRS elements with the fixed distributed AIRS scheme.

\subsubsection{User-Adaptive AIRS Beamforming}
\begin{figure}[t]
	\centering
	\includegraphics[width=0.95\linewidth]{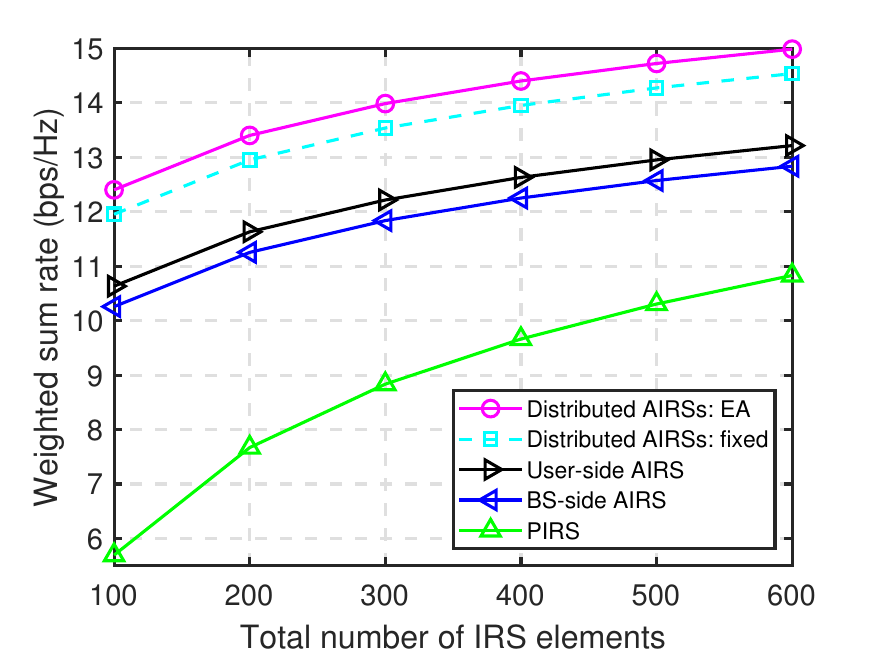}
	\caption{WSR versus total number of IRS elements $N$.}
	\label{fig:UA}
\end{figure}

As a comparison, we consider the following schemes: 1) \textbf{Distributed AIRS: EA}: the results are obtained by our proposed element allocation design; 4) \textbf{Distributed AIRS: fixed}: the element allocation is fixed at the distributed AIRSs. Fig. \ref{fig:UA} shows the WSR versus the total number of IRS elements. One can observe that the proposed design for distributed AIRS achieves superior performance among all considered schemes. To achieve a sum rate of 13 bps/Hz, deploying distributed AIRSs reduces the required number of elements from 600 to 200 compared to the BS/user-side AIRS. This improvement stems from strategic deployment, which compensates for performance loss due to fewer active elements and balances the trade-off between signal and noise amplification. Notably, even deploying the BS-side AIRS can achieve significant gains over employing PIRS in terms of the system sum rate thanks to its amplification capability, which mitigates the severe double path loss while enhancing received signal power. In the following, we focus on the WSR maximization problem by optimizing the element allocation at the distributed AIRS.

\begin{figure}[t]
	\centering
	\includegraphics[width=0.95\linewidth]{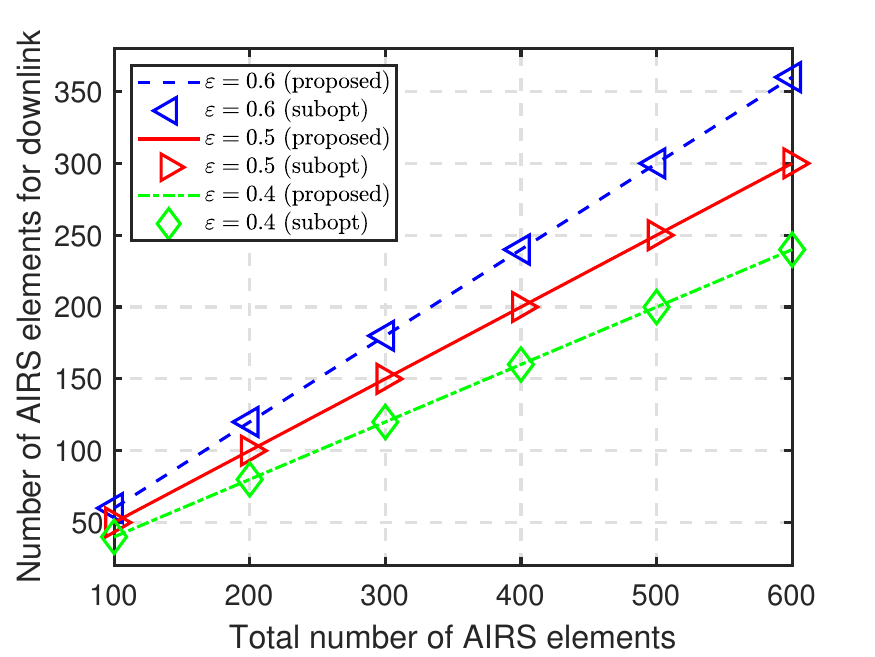}
	\caption{Number of AIRS elements for downlink $N_\mathrm{D}$ versus total number of AIRS elements $N$ under the multi-user setup.}
	\label{fig:UA_t}
\end{figure}
In Fig. \eqref{fig:UA_t}, we plot the number of AIRS elements for DL $N_\mathrm{D}$ versus total number of AIRS elements $N$ for $\varepsilon = 0.4$, $\varepsilon = 0.5$, and $\varepsilon = 0.6$. One can observe that the number of AIRS elements for DL monotonically increases as $N$ increase. In addition, for a fixed $N$, $N_\mathrm{D}$ increases with $\varepsilon$. Moreover, it is observed that the sub-optimal solution, i.e., $x^{\mathrm{subopt}}_\mathrm{D} = \varepsilon N$, is close to the solution obtained by the proposed exhaustive search. This is because the results are obtained in the high SNR case of both UL and DL communications. It is consistent with the discussion in Section \ref{sec:user-adaptive}, which validates the effectiveness of such a simple element allocation strategy.

\subsubsection{Static/Constant AIRS Beamforming}
\begin{figure}[t]
	\centering
	\includegraphics[width=0.95\linewidth]{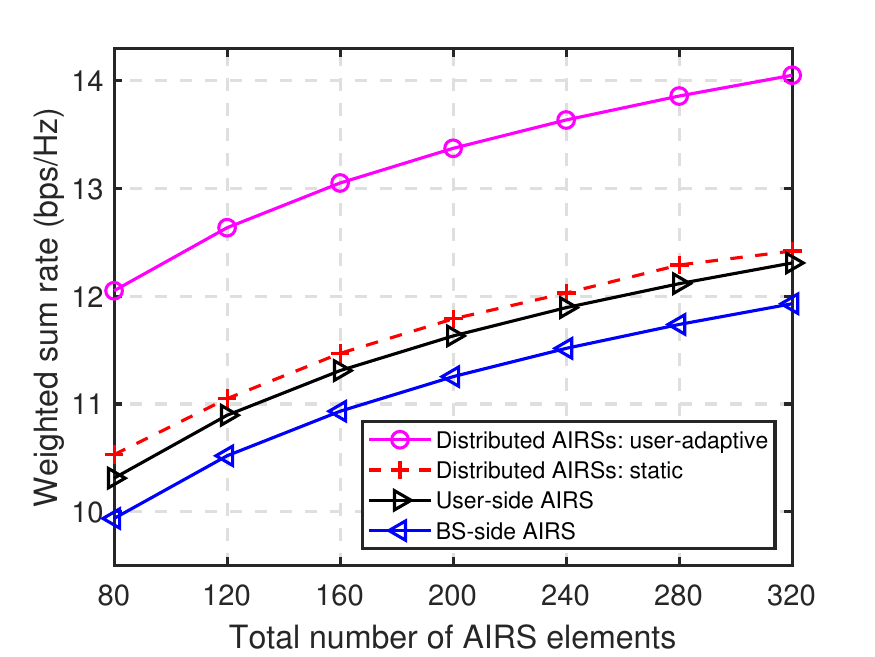}
	\caption{WSR versus total number of AIRS elements $N$.}
	\label{fig:static}
\end{figure}
For comparison, we consider the following schemes with AIRS: 1) \textbf{Distributed AIRS: user-adaptive}: the user-adaptive AIRS beamforming is adopted with equal element allocation; 2) \textbf{Distributed AIRS: static}:  Algorithm \ref{Algo: 1} is employed to obtain the WSR under the static AIRS beamforming configuration. Fig. \ref{fig:static} shows the rate performance comparison between the considered schemes by plotting the WSR versus the total number of AIRS elements when $\varepsilon = 0.4$. It shows that employing the user-adaptive beamforming at the distributed AIRSs can achieves a significant gain compared with static beamforming, which results from the following two reasons. First, due to the different locations of the AIRSs and the asymmetry of the UL and DL channels, employing identical phase configurations leads to beam misalignment and degraded signal quality. Second, in multi-user scenarios, fixed amplification factors cannot adapt to the channel differences among different users. Nevertheless, the static beamforming scheme offers notable practical advantages, such as easy-implementation and no need for frequent beamforming optimization, thereby significantly reducing computational overhead and channel feedback requirements, which makes it particularly suitable for scenarios where low latency is not critical. Interestingly, the distributed AIRS with static beamforming can outperform the BS/user-side AIRS with dynamic beamforming, which further emphasizes the importance of the deployment architecture. The results highlight the important role of deployment architecture in AIRS-aided joint UL and DL communication systems and suggest that strategic AIRS placement can compensate for the limitations of static beamforming scheme.

\begin{figure}[t]
	\centering
	\includegraphics[width=0.95\linewidth]{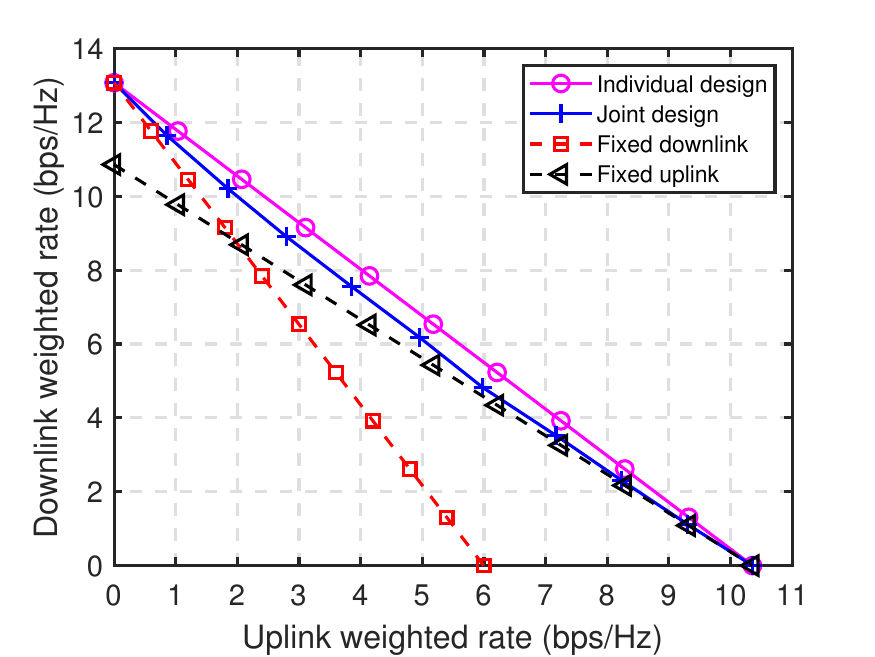}
	\caption{Uplink-downlink rate region for $N=100$.}
	\label{fig:ULDL}
\end{figure}
To explore the UL-DL rate region, problem \eqref{pro:tdma} is optimized for different values of $\varepsilon$ from 0 to 1. For comparison, four weighting schemes are considered: 1) \textbf{Individual design}: the AIRS beamforming for UL and DL transmissions is optimized individually; 2) \textbf{Joint design}: the WSR is maximized via Algorithm \ref{Algo: 1}; 3) \textbf{Fixed downlink}: the DL weighted rate is maximized by jointly optimizing $\left\{ {{{\bm{w}}_k}} \right\}$, ${{\bm{\Phi }}}$, and ${\alpha _{{\mathrm{D}}}}$. With optimized ${{\bm{\Phi }}}$, the WSR is maximized by jointly optimizing $\left\{ {{{\bm{u}}_k}} \right\}$ and ${\alpha _{{\mathrm{U}}}}$; 4) \textbf{Fixed uplink}: the UL weighted rate is maximized by jointly optimizing $\left\{ {{{\bm{u}}_k}} \right\}$, ${{\bm{\Phi }}}$, and ${\alpha _{{\mathrm{U}}}}$. With optimized ${{\bm{\Phi }}}$, the WSR is maximized by jointly optimizing $\left\{ {{{\bm{w}}_k}} \right\}$ and ${\alpha _{{\mathrm{D}}}}$. Fig. \ref{fig:ULDL} shows the trade-off between UL weighted rate versus DL weighted rate. Compared with the fixed UL and fixed DL schemes, the joint design achieves a larger rate region, thereby enhancing the overall system performance. This is because fixed design schemes maximize the rate in a single transmission link at the expense of the rate in the other link, leading to a highly unbalanced system performance. In contrast, the joint design enables a more flexible trade-off between UL and DL rates, allowing the system to satisfy diverse service requirements. The corresponding curve bends inward and shows nonlinearity, which highlights the complex relationship between resource allocation and system performance and the asymmetry between UL and DL. It is also worth noting that achieving the individual design boundary typically requires dynamic beamforming configurations, which may introduce additional hardware complexity and signaling overhead. Considering both performance and efficiency, the joint design is a more practical solution for deployment.

\section{Conclusion}
\label{Conclusion}
In this paper, we investigated the WSR of the distributed AIRS-aided joint UL and DL communications. In the single-user case, we provided an analytical framework to theoretically compare the WSR achieved by the distributed, BS-side, and user-side AIRSs. In the multi-user scenario, we studied the AIRS element allocation optimization problem with the user-adaptive AIRS beamforming design for sum rate maximization. Furthermore, to reduce the signaling overhead, we adapted an efficient static AIRS beamforming scheme for the distributed AIRS. The AO-based algorithm was developed to address the WSR maximization problem by jointly optimizing the BS, user, and AIRS beamforming. Numerical results validated our theoretical findings and demonstrated the superiority of the distributed AIRS in terms of maximizing the WSR, supporting multiple users, and reducing the required number of AIRS elements. Although static AIRS beamforming may experience performance degradation, strategic deployment of distributed AIRSs allows them to outperform centralized dynamic schemes while maintaining low computational overhead.

\section*{Appendix A}
\section*{Proof of Proposition \ref{pro:5}}
For any given ${\left\{ {p_k} \right\}}$, ${\left\{ {{{\bm{w}}_k}} \right\}}$, ${\left\{ {{{\bm{u}}_k}} \right\}}$, ${{\bm{\Phi }}}$, and ${\alpha _{{\mathrm{D}}}}$, constraint \eqref{a_ul} is strictly satisfied with equality because the objective value can be improved by increasing $\alpha _{{\mathrm{U}}}$ until constraint \eqref{a_ul} is active. Thus, the optimal amplification factor at the BS-side AIRS is given by
\begin{align}
	{\alpha _{{\mathrm{U}}}^\mathrm{opt}} = \min \left\{{\sqrt {\frac{{{P_{\mathrm{F}}}}}{{{{p_k}{\| {{\bm{\Phi h}}_{\mathrm{U},k}} \|^2} + \sigma _{\mathrm{F}}^2 N } }}} }, \forall k \in \mathcal{K} \right\}.
\end{align}
For any given ${\left\{ {{{\bm{w}}_k}} \right\}}$, ${\left\{ {{{\bm{u}}_k}} \right\}}$, ${{\bm{\Phi }}}$, ${\alpha _{{\mathrm{U}}}}$, and ${\alpha _{{\mathrm{D}}}}$, problem \eqref{pro:tdma} is reduced to 
\begin{subequations}
	\label{pro:p_1}
	\begin{align}
		\mathop {\max }\limits_{\left\{ {{p_k}} \right\}} \; &\sum \nolimits_{k = 1}^K {R_{\mathrm{U},k}} \label{pro:p_obj}\\
		\mathrm{s.t.} \;\;
		& 0 \le {p_k} \le {P_{\mathrm{U}}},\forall k \in \mathcal{K}, \\
		&\alpha _{{\mathrm{U}}}^2( {{p_k}{\| {{\bm{\Phi h}}_{\mathrm{U},k}} \|^2} + \sigma _{\mathrm{F}}^2\left\| {\bm{\Phi }} \right\|_F^2} ) \le {P_{\mathrm{F}}}, \forall k \in \mathcal{K}.
	\end{align}
\end{subequations}
In the following, we prove $p_k^\mathrm{opt} = P_\mathrm{U}, \forall k \in \mathcal{K}$ by contradiction. Assume that $\left\{p_k^\mathrm{opt},{\alpha _{{\mathrm{U}}}^\mathrm{opt}}\right\}$ is the optimal transmit power of user $k$ and amplification vector and $p_k^\mathrm{opt} < P_\mathrm{U}$. When $p_k^\mathrm{opt}{\| {{\bm{\Phi h}}_{\mathrm{U},k}} \|^2} = \max \{{p_i}{\| {{\bm{\Phi h}}_{\mathrm{U},i}} \|^2}, \forall i \in \mathcal{K}\}$ and ${\alpha _{{\mathrm{U}}}^\mathrm{opt}} = \sqrt {\frac{{{P_{\mathrm{F}}}}}{{{{p_k^\mathrm{opt}}{\| {{\bm{\Phi h}}_{\mathrm{U},k}} \|^2} + \sigma _{\mathrm{F}}^2 N } }}} $, we can construct a solution denoted by $\left\{\tilde p_k, \tilde \alpha _{\mathrm{U}}\right\}$, which satisfies $p_k^\mathrm{opt} < \tilde p_k \le P_\mathrm{U}$ and $\tilde \alpha _{\mathrm{U}} = \sqrt{p_k^\mathrm{opt}/\tilde p_k} {\alpha _{{\mathrm{U}}}^\mathrm{opt}}$. With $\tilde \alpha _{{\mathrm{U}}}^2( {{\tilde p_k}{\| {{\bm{\Phi h}}_{\mathrm{U},k}} \|^2} + \sigma _{\mathrm{F}}^2\left\| {\bm{\Phi }} \right\|_F^2} ) < (\alpha _{{\mathrm{U}}}^\mathrm{opt})^2( {{p_k^\mathrm{opt}}{\| {{\bm{\Phi h}}_{\mathrm{U},k}} \|^2} + \sigma _{\mathrm{F}}^2\left\| {\bm{\Phi }} \right\|_F^2} ) \le {P_{\mathrm{F}}}$, it indicates that $\left\{\tilde p_k, \tilde \alpha _{\mathrm{U}}\right\}$ is feasible for problem \eqref{pro:p_1}. By comparing the rate with the two solutions, we have 
\begin{align}
	\label{solution_com_1}
	&\frac{1}{K}{\log _2} \left( \! 1 \!+\! \frac{{{p_k^\mathrm{opt}}{| {{\bm{u}}_k^H{\bm{G}}_{{\mathrm{U}}}^H{\alpha _{{\mathrm{U}}}^\mathrm{opt}}{{\bm{\Phi }}}{\bm{h}}_{\mathrm{U},k}} |^2}}}{{{\bm{u}}_k^H ( {(\alpha _{{\mathrm{U}}}^\mathrm{opt})^2\sigma _{\mathrm{F}}^2{\bm{G}}_{{\mathrm{U}}}^H{{\bm{\Phi }}}{\bm{\Phi }}^H{{\bm{G}}_{{\mathrm{U}}}} \!+\! \sigma _0^2{{\bm{I}}_M}} ) {{\bm{u}}_k}}} \right) \nonumber\\
	&< \! \frac{1}{K}{\log _2}  \left( \! 1 \!+\! \frac{{{\tilde p_k}{| {{\bm{u}}_k^H{\bm{G}}_{{\mathrm{U}}}^H{\tilde \alpha _{{\mathrm{U}}}}{{\bm{\Phi }}}{\bm{h}}_{\mathrm{U},k}} |^2}}}{{{\bm{u}}_k^H ( {\tilde \alpha _{{\mathrm{U}}}^2\sigma _{\mathrm{F}}^2{\bm{G}}_{{\mathrm{U}}}^H{{\bm{\Phi }}}{\bm{\Phi }}^H{{\bm{G}}_{{\mathrm{U}}}} \!+\! \sigma _0^2{{\bm{I}}_M}} ) {{\bm{u}}_k}}} \right).
\end{align}
Inequality \eqref{solution_com_1} indicates that the constructed solution $\left\{\tilde p_k, \tilde \alpha _{\mathrm{U}}\right\}$ yields a higher WSR, which conflicts with the claim that $\left\{p_k^\mathrm{opt},{\alpha _{{\mathrm{U}}}^\mathrm{opt}}\right\}$ is optimal. When $p_k^\mathrm{opt}{\| {{\bm{\Phi h}}_{\mathrm{U},k}} \|^2} < \max \{{p_i}{\| {{\bm{\Phi h}}_{\mathrm{U},i}} \|^2}, \forall i \in \mathcal{K}\} = p_m{\| {{\bm{\Phi h}}_{\mathrm{U},m}} \|^2}$ and ${\alpha _{{\mathrm{U}}}^\mathrm{opt}} = \sqrt {\frac{{{P_{\mathrm{F}}}}}{{{{p_m}{\| {{\bm{\Phi h}}_{\mathrm{U},m}} \|^2} + \sigma _{\mathrm{F}}^2 N } }}} $, we can construct a solution denoted by $\bar p_k$, which satisfies $p_k^\mathrm{opt} < \bar p_k \le p_m = P_\mathrm{U}$. With $(\alpha _{{\mathrm{U}}}^\mathrm{opt})^2 ( {{\bar p_k}{\| {{\bm{\Phi h}}_{\mathrm{U},k}} \|^2} + \sigma _{\mathrm{F}}^2\left\| {\bm{\Phi }} \right\|_F^2} ) < (\alpha _{{\mathrm{U}}}^\mathrm{opt})^2( {{p_m}{\| {{\bm{\Phi h}}_{\mathrm{U},m}} \|^2} + \sigma _{\mathrm{F}}^2\left\| {\bm{\Phi }} \right\|_F^2} ) \le {P_{\mathrm{F}}}$, it indicates that $\bar p_k$ is feasible for problem \eqref{pro:p_1}. By comparing the rate with the two solutions, we have 
\begin{align}
	\label{solution_com_2}
	&\frac{1}{K}{\log _2} \!\! \left( \!\! 1 \!\!+\!\! \frac{{{p_k^\mathrm{opt}}{| {{\bm{u}}_k^H{\bm{G}}_{{\mathrm{U}}}^H{\alpha _{{\mathrm{U}}}^\mathrm{opt}}{{\bm{\Phi }}}{\bm{h}}_{\mathrm{U},k}} |^2}}}{{{\bm{u}}_k^H ( {(\alpha _{{\mathrm{U}}}^\mathrm{opt})^2\sigma _{\mathrm{F}}^2{\bm{G}}_{{\mathrm{U}}}^H{{\bm{\Phi }}}{\bm{\Phi }}^H{{\bm{G}}_{{\mathrm{U}}}} \!\!+\! \sigma _0^2{{\bm{I}}_M}} ) {{\bm{u}}_k}}} \!\! \right) \nonumber\\
	&< \!\! \frac{1}{K}{\log _2} \!\! \left( \!\! 1 \!\!+\!\! \frac{{{\bar p_k}{| {{\bm{u}}_k^H{\bm{G}}_{{\mathrm{U}}}^H{\alpha _{{\mathrm{U}}}^\mathrm{opt}}{{\bm{\Phi }}}{\bm{h}}_{\mathrm{U},k}} |^2}}}{{{\bm{u}}_k^H (\! {(\alpha _{{\mathrm{U}}}^\mathrm{opt})^2\sigma _{\mathrm{F}}^2{\bm{G}}_{{\mathrm{U}}}^H{{\bm{\Phi }}}{\bm{\Phi }}^H{{\bm{G}}_{{\mathrm{U}}}} \!\!+\! \sigma _0^2{{\bm{I}}_M}} \!) {{\bm{u}}_k}}} \!\! \right).
\end{align}
Inequality \eqref{solution_com_2} indicates that the constructed solution $\left\{\tilde p_k, \tilde \alpha _{\mathrm{U}}\right\}$ yields a higher WSR, which conflicts with the claim that $p_k^\mathrm{opt}$ is optimal. Thus, the proof is completed.

\bibliographystyle{IEEEtran}
\bibliography{refs.bib} 

\end{document}